\documentclass[lettersize,journal]{IEEEtran}
\usepackage{array}
\usepackage{textcomp}
\usepackage{stfloats}
\usepackage{url}
\usepackage{verbatim}
\usepackage{graphicx}
\usepackage{cite}
\usepackage{amsmath,amssymb,amsfonts}
\usepackage{xcolor}
\usepackage{bm}
\usepackage{acronym}
\usepackage{cancel}
\usepackage{balance}
\usepackage[ruled,vlined,linesnumbered]{algorithm2e}
\usepackage{tikz}
\usepackage{subcaption}
\usepackage{adjustbox}   
\usepackage{comment}
\usepackage{booktabs}
\usepackage[table]{xcolor}
\usepackage{hyperref}
\newtheorem{assumption}{Assumption}

\def\BibTeX{{\rm B\kern-.05em{\sc i\kern-.025em b}\kern-.08em
    T\kern-.1667em\lower.7ex\hbox{E}\kern-.125emX}}

    \makeatletter
\def\ps@IEEEtitlepagestyle{%
    \def\@oddhead{\footnotesize\hfill
    This work has been submitted to IEEE for possible publication.
    Copyright may be transferred without notice, after which this
    version may no longer be accessible.\hfill}
    \def\@evenhead{\@oddhead}
    \def\@oddfoot{\hfill\thepage\hfill}
    \def\@evenfoot{\hfill\thepage\hfill}
}
\makeatother

\begin{document}

\title{An Encoded Corrective Double Deep Q-Networks for Multi-Agent Control Systems
\thanks{The authors are with University of California, Irvine
CA 92697, USA (e-mail: \{barzegm1, kemengh, hamidj\}@uci.edu). This work was supported in part by ARO grant W911NF2410046.}
}

\author{\IEEEauthorblockN{Mohammadreza Barzegaran, Kemeng Han, and Hamid Jafarkhani}
}

\maketitle

\acrodef{lti}[LTI]{linear time-invariant}
\acrodef{mas}[MAS]{multi-agent system}
\acrodef{dqn}[DQN]{Deep Q-Network}
\acrodef{coco}[CDNet]{encoded corrective double deep Q-networks}
\acrodef{uav}[UAV]{uncrewed aerial vehicle}
\acrodef{marl}[MARL]{Multi-Agent Reinforcement Learning}
\acrodef{e2e}[E2E]{end-to-end}
\acrodef{nawgn}[NAWGN]{non-zero mean additive white Gaussian noise}
\acrodef{ema}[EMA]{exponential moving average}

\begin{abstract}
This paper studies the synthesis of control policies for heterogeneous and interconnected multi-agent systems that collaborate through data exchange over a communication network to minimize a collective cost. We propose a distributed encoded corrective double actor-critic framework that integrates a novel message-passing mechanism. Existing methods assume noise-free and delay-free access to the global or partial states and overlook the fact that the global states, though noisy and delayed, can be progressively reconstructed and refined over time. In contrast, this work explicitly models communication sampling asynchrony, delay, and link noise based on the network configuration. The proposed message-passing mechanism characterizes timing and information flow to refine and time shift global state information, which is then used to incrementally correct the Q-networks. The double Q-network design mitigates overestimation bias, while the shared encoder coupling the actor-critic networks captures inter-agent dependencies. We evaluate our approach in multiple test cases, demonstrate its effectiveness over various baselines, and provide a numerical regret analysis.
\end{abstract}

\begin{IEEEkeywords}
Multi-agent systems, Delayed networks, Asynchronous communication, Noisy observation, Deep reinforcement learning
\end{IEEEkeywords}

\section{Introduction}
A \ac{mas} refers to a collection of individual agents that sense, interact, and coordinate within a shared environment to achieve a common objective~\cite{whatisMAS}. In distributed control, \acp{mas} are widely used because of their ability to independently learn local control policies for each agent while collectively minimizing a global objective~\cite{featureMAS}. In this context, most \acp{mas} are interconnected through (\emph{i}) physical couplings, where the state of an agent affects others (e.g., collision avoidance),  (\emph{ii}) a shared objective that requires cooperation, or (\emph{iii}) a combination of both~\cite{alemzadeh2019distributed,wang2020distributed,d2003distributed}. Synthesizing control policies for such systems is more challenging, as each agent must account for the dynamic influence of others while estimating the global state.

To this end, various methods have been proposed under the assumption of known system models, including decomposition-based approaches~\cite{massioni2009distributed,hoffmann2013distributed}, model predictive control~\cite{venkat2005stability}, receding horizon control~\cite{dunbar2007distributed}, and LQR consensus control~\cite{li2015fully,chang2023regret}, among others. However, in large-scale interconnected \acp{mas}, the system model is often unavailable, rendering analytical solutions impractical. Even when analytical approaches are extended to handle unknown systems, they typically rely on unrealistic assumptions, such as agent homogeneity~\cite{borrelli2008distributed,vlahakis2019distributed} or fully connected agent networks~\cite{dong2010distributed}.

Nevertheless, model-free reinforcement learning has demonstrated strong potential for synthesis of control in such systems. This line of work began with the seminal work in~\cite{watkins1992q}, which introduced model-free Q-leaning to single-agent systems, and was later extended to centralized control synthesis for \acp{mas} with unknown system models~\cite{narayanan2016distributed,dizche2019sparse}. However, these centralized decision-making frameworks are often impractical for large-scale systems and infeasible in many real-world applications. Consequently, distributed reinforcement learning approaches that rely on information exchange among agents over communication networks have emerged. For example, in \cite{zhang2018fully, zhang2019distributed,ma2024efficient}, agents exchange their local estimates to achieve global consensus and collaboratively learn control policies. In \cite{kayaalp2023policy}, a belief-sharing algorithm is developed to diffuse information and estimate the global observation. Similarly, \cite{sharedmemory2020,offlinebeliefgeneration2021} propose learning compact representations of each agent's local history to estimate global belief in partially observed environments. In~\cite{chang2023regret}, a fully distributed online learning algorithm is proposed for consensus in multi-agent linear systems with unknown networks. Finally, \cite{wang2020distributed, he2019adaptive} introduce policy iteration-based methods to synthesize control for \acp{mas} with unknown system models.

Although many existing methods rely on communication networks for coordination among agents, they often make ideal assumptions and overlook key network-induced complexities including link noise, communication delays, and asynchrony. Classical consensus studies explicitly characterize the impact of these imperfections on convergence and stability~\cite{aysal2008distributed,kar2008distributed,rajagopal2010network,griparic2022consensus}. Building upon these works, subsequent studies in distributed learning~\cite{ortega2024quantized,ortega2024decentralized,shen2021distributed,diaz2025multi}, control~\cite{diaz2025multi,barzegaran2025dynamic,tung2021effective,yang2023collision}, and deployment~\cite{9930941,10186347}  incorporate communication imperfections and constraints into decision making, and analyze their impact on achievable performance, stability, and convergence. However, under such network-induced imperfections, accurate, immediate, and complete state information is generally unavailable to individual agents. Moreover, while many approaches account for various communication imperfections, they fail to exploit the network topology to progressively estimate and refine the global state over time. In contrast, this work leverages the structure of the communication graph to model timing and information flow, enabling the mitigation of asynchrony, delay, and link noise through network-aware state reconstruction and refinement.

In this work, we address the problem of distributed control in \ac{mas} networks subject to noisy, asynchronous sampling, and delayed communication. Specifically, our goal is to synthesize linear feedback control policies for each agent in a distributed manner to minimize the collective quadratic cost of a \ac{lti} \ac{mas}. The use of \ac{lti} dynamics and a collective quadratic cost provides a ground truth for convergence analysis; however, the proposed approach is general and applicable to non-\ac{lti} systems as well. Due to the interconnection among agents, the global system model is unknown and cannot be expressed analytically. Agents establish and maintain a time-invariant communication network through which they exchange data for cooperative policy learning.

We propose \ac{coco} algorithm which implements a distributed actor-critic framework. The actor networks determine control policies that minimize the collective cost. To mitigate overestimation bias in the value estimation of critic networks~\cite{overestimate}, which is amplified by noise, we employ double critic Q-networks and adopt pessimistic Q-value updates through target updates. Furthermore, the actor-critic networks are coupled through a shared encoder that captures both agent dynamics and inter-agent dependencies. The encoder processes a window of observations to extract temporal features.

To address asynchronous sampling, delayed, and noisy communication in \ac{mas} networks, \ac{coco} implements a message-passing mechanism which leverages the network graph. This mechanism identifies communication routes between pair of agents and simulates sampled state estimation over the network. For each route, the number of hops is computed to determine relative communication timing, which corresponds to discrete sample-delay steps. Asynchrony is compensated using this hop-based delay model. In parallel, the cumulative link noise along each route is estimated and used to refine the received information.
%compute both the number of hops and the cumulative channel noise along each route. Asynchrony is compensated using the hop count to determine the relative communication timing, while the received information is refined according to the estimated cumulative noise of the link over the corresponding route.
During learning, each agent first updates its global state estimation using immediately available noise-mitigated data. Subsequently, as delayed information becomes available according to the relative timing, the agent performs soft corrections to its state estimates. To further enhance robustness, the message-passing mechanism employs a Dijkstra-based route selection algorithm that balances communication delay and cumulative noise, enabling more accurate and consistent reconstruction of the global state.

The contributions of the paper are as follows:
\begin{itemize}
    \item We propose a message-passing mechanism that simulates sampled state estimation and models communication sampling asynchrony, delay, and link noise based on the network graph. The mechanism uses hop counts and link noise statistics to compensate relative timing and refine received observations.  
    \item We develop a Dijkstra-based route selection algorithm that balances communication delay and cumulative link noise, ensuring a more accurate and consistent reconstruction of the global state.
    \item We design a distributed actor–critic framework in which double critic networks mitigate overestimation bias amplified by noise through pessimistic Q-value updates.
    \item To capture inter-agent dependencies, the actor and critic networks are coupled through a shared encoder that extracts temporal features from historical state estimation information, improving robustness under noisy, asynchronous, and delayed communication.
%    \item We provide extensive evaluation, including empirical convergence analysis, establish the convergence properties of our framework under an asynchronous setting, and compare with the state-of-the-art. 
\end{itemize}
We also provide extensive evaluation, including numerical regret analysis, establish the convergence properties of our framework under an asynchronous setting, and compare it with the state-of-the-art. 

The organization of the paper is as follows. Section~\ref{sec:motivation} presents an illustrative example to motivate the problem in hand. Section~\ref{sec:preliminaries} presents the preliminaries of control in \acp{mas}. Section~\ref{sec:messagepassing} discusses the message-passing mechanism and how agents exchange information in networks. Section~\ref{sec:proposed} presents our novel \ac{coco} algorithm. We evaluate \ac{coco} on various test cases and provide an analysis of their effectiveness in Section~\ref{sec:evaluation}. Concluding remarks and future work are discussed in Section~\ref{sec:conclusion}.

\section{Motivation}
\label{sec:motivation}

\begin{figure}
\centering
\includegraphics[width=0.95\linewidth]{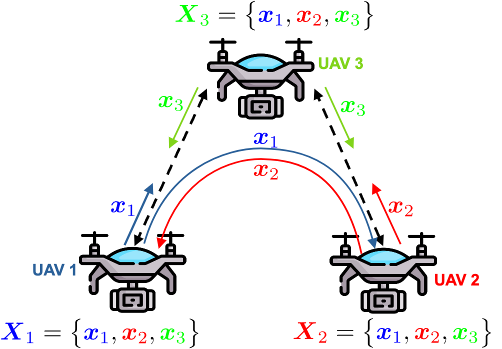}
\caption{Illustrative example: UAVs exchange data to build the global states information and compute control, assuming ideal communication without delays and noise.}
\label{fig:motivation-nodelay}
\end{figure}

This section presents an illustrative example to motivate the problem at hand. Consider a group of three \acp{uav} operating in a shared environment to perform a collaborative task, such as formation flight. The objective is to control each \ac{uav} such that the group collectively minimizes a common cost function. Each \ac{uav} shares its state (e.g., position) with others through a multi-hop communication network, allowing all \acp{uav} to access the global state information required for control via multi-hop data forwarding. A hop refers to a single communication link between two directly connected \acp{uav}; data transmitted from one \ac{uav} to another over multiple intermediate \acp{uav} is said to traverse multiple hops.
%In such networks, data is transmitted via packets, each includes the ID of the sender, the state information, and the relative time stamp generated by the sender.

As an example, in Fig.~\ref{fig:motivation-nodelay}, \acp{uav}~1 and~3 share their state information with \ac{uav}~2, denoted by $\bm{x}_1$ and $\bm{x}_3$, respectively. Using these data, \ac{uav}~2 builds the global state $\bm{X}=\big\{\bm{x}_1,\bm{x}_2,\bm{x}_3\big\}$ for control computation. In the same way, \ac{uav}~1 receives $\bm{x}_2$ and $\bm{x}_3$, and \ac{uav}~3 receives $\bm{x}_1$ and $\bm{x}_2$. This allows every \ac{uav} to reach an agreement on a common global state information and calculate its control action based on this information. However, this figure overlooks the complexities of communication networks and assumes continuous, synchronous sampling, delay-free, and noise-free communication.

\begin{figure*}[!t]
  \centering
  \begin{subfigure}{0.49\linewidth}
    \includegraphics[width=0.95\linewidth]{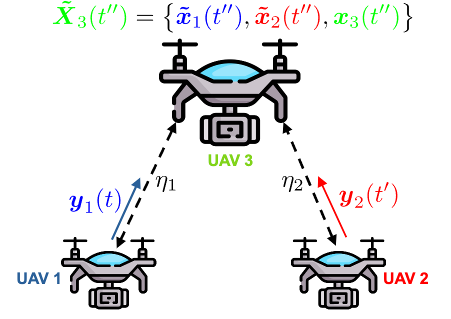}
    \caption{Network communication at $t^{\prime\prime}$: UAV~3 infers that the local timestamps are effectively synchronized, i.e., $t = t^\prime = t^{\prime\prime}$, and constructs the refined global states information.}
    \label{fig:motivation-delay-case1}
  \end{subfigure}
  \hfill
  \begin{subfigure}{0.49\linewidth}
    \includegraphics[width=\linewidth]{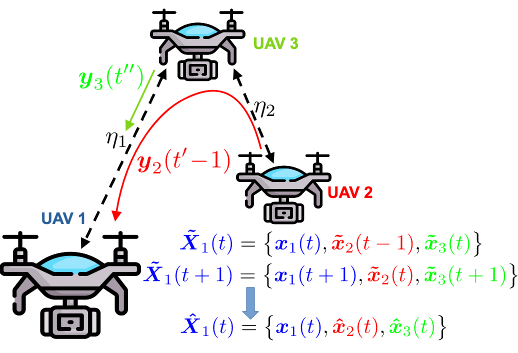}
    \caption{Network communication at $t$ and $t+1$: UAV~1 uses the effective synchronization $t = t^\prime = t^{\prime\prime}+1$ to reconstruct the global states estimation.}
    \label{fig:motivation-delay-case2}
  \end{subfigure}
  \caption{Illustrative example: Information exchange in UAV networks are subject to noise, asynchrony, and delay.}
  \label{fig:motivation-delay}
\end{figure*}

Fig.~\ref{fig:motivation-delay} illustrates a realistic communication scenario in which information is sampled asynchronously and is subject to communication delays and noise. In this setting, \acp{uav} operate on sampled data defined over discrete time domain, i.e., $x(t), x(t+1), \ldots$. Each \ac{uav} relies on its own reference time. Because these local clocks are not necessarily synchronized, information sampling occurs asynchronously.
Let the local timestamps of \acp{uav}~1, 2, and 3 be denoted by $t$, $t^\prime$, and $t^{\prime\prime}$, respectively. For simplicity and without loss of generality, we assume that the data propagate instantaneously over each link and are forwarded in one discrete time step. Under this assumption, a two-hop transmission introduces a one-step delay. Consequently, \ac{uav}~3 receives and transmits information instantaneously, whereas \acp{uav}~1 and~2 receive each other’s information with a one-step delay while receiving information from \ac{uav}~3 instantaneously.

In addition to the delay, each link introduces noise as well. As a result, the received information is affected by the accumulation of link noise along the communication route.
%This end-to-end noise is applied to the received information.
We denote the link noise between \acp{uav}~1 and 3 as $\eta_1$, and that between \acp{uav}~2 and 3 as $\eta_2$. Fig.~\ref{fig:motivation-delay-case1} shows the network communication observed at \ac{uav}~3's local timestamp $t^{\prime\prime}$. As expected, this \ac{uav} instantaneously receives the noisy state information from \acp{uav}~1 and~2 as $\bm{y}_1(t) = \bm{x}_1(t)+\eta_1$ and $\bm{y}_2(t^\prime) = \bm{x}_2(t^\prime)+\eta_2$, respectively. Note that $\bm{x}_3(t^{\prime\prime})$ is not subject to noise. Therefore, \ac{uav}~3's observation of the global state is 
\begin{align*}
    \bm{Y}_3(t^{\prime\prime})=\big\{\bm{y}_1(t^{\prime\prime}),\bm{y}_2(t^{\prime\prime}),\bm{x}_3(t^{\prime\prime})\big\}.
\end{align*}
Then, \ac{uav}~3 constructs its refined global state as
\begin{align*}
    \bm{\tilde{X}}_3(t^{\prime\prime})=\big\{\bm{\tilde{x}}_1(t^{\prime\prime}),\bm{\tilde{x}}_2(t^{\prime\prime}),\bm{x}_3(t^{\prime\prime})\big\},
\end{align*}
where the operator $\tilde{}$ denotes the refined/filtered version of the state information. Since \ac{uav}~3 experiences instantaneous communication with both neighboring agents, it infers that the local timestamps are effectively synchronized, i.e., $t = t^\prime = t^{\prime\prime}$. Therefore, the constructed global state can then be directly used for control purposes.

Fig.~\ref{fig:motivation-delay-case2} shows the network communication received by \ac{uav}~1. As expected, this \ac{uav} instantaneously receives the noisy state information from \ac{uav}~3, while the noisy information from \ac{uav}~2 arrives with a one-step delay. At \ac{uav}~1's local timestamp $t$, it receives $\bm{y}_3(t^{\prime\prime}) = \bm{x}_3(t^{\prime\prime})+\eta_1$ and $\bm{y}_2(t^{\prime}-1) = \bm{x}_2(t^{\prime}-1)+\eta_1 + \eta_2$ from \acp{uav}~3 and 2, respectively. Similarly, at the next local timestamp $t+1$, it receives $\bm{y}_3(t^{\prime\prime}+1) = \bm{x}_3(t^{\prime\prime}+1)+\eta_1$ and $\bm{y}_2(t^{\prime}) = \bm{x}_2(t^{\prime})+\eta_1 + \eta_2$. Note that $\bm{x}_1(t)$ and $\bm{x}_1(t+1)$ are \ac{uav}~1's real state information, not affected by delay or noise. Similarly, \ac{uav}~1 constructs its refined global state as
\begin{align*}
    \bm{\tilde{X}}_1(t)=&\big\{\bm{x}_1(t),\bm{\tilde{x}}_2(t-1),\bm{\tilde{x}}_3(t)\big\},\\
    \bm{\tilde{X}}_1(t+1)=&\big\{\bm{x}_1(t+1),\bm{\tilde{x}}_2(t),\bm{\tilde{x}}_3(t+1)\big\},
\end{align*}
at local timestamps $t$ and $t+1$, respectively.
\ac{uav}~1 uses the refined global state~$\bm{\tilde{X}}_1$ to learn its control policy. Then, continuously, \ac{uav}~1 aligns timestamps of $\bm{\tilde{x}}_2$ using the effective synchronization $t = t^\prime = t^{\prime\prime}+1$ to reconstruct the global state estimation as
\begin{align*}
    \bm{\hat{X}}_1(t)=&\big\{\bm{x}_1(t),\bm{\hat{x}}_2(t),\bm{\hat{x}}_3(t)\big\}.
\end{align*}
This reconstructed information is later used to refine the learning process. The same reasoning applies to \ac{uav}~2.

The proposed message-passing mechanism performs two key functions: (\emph{i}) it refines and filters noisy information, and (\emph{ii}) it determines the relative timestamp synchronization to reconstruct the global state estimation accordingly. This mechanism leverages graph-based analysis for information estimation, synchronization, and refinement.

\section{Preliminaries}
\label{sec:preliminaries}
{\em Notation:} Lowercase letters ($a$, $b$, $\ldots$) and uppercase letters ($A$, $B$, $\ldots$) denote scalars. Bold lowercase letters ($\bm{a}$, $\bm{b}$, $\ldots$) and bold uppercase letters ($\bm{A}$, $\bm{B}$, $\ldots$) represent vectors and matrices, respectively. Blackboard bold uppercase letters ($\mathbb{A}$, $\mathbb{B}$, $\ldots$) denote sets, where the corresponding regular uppercase letters indicate their cardinalities. Function mappings, including Q-networks as function approximators, are denoted by lowercase and uppercase letters (e.g., $f(\cdot)$ and $F(\cdot)$) for scalar-valued functions, and bold lowercase letters (e.g., $\bm{f}(\cdot)$) for vector-valued functions. Calligraphic uppercase letters ($\mathcal{A}$, $\mathcal{B}$, $\ldots$) represent block matrices. A column vector is defined as $\bm{X} = \mathrm{col}(\bm{x}_1, \bm{x}_2, \ldots, \bm{x}_L)$, stacking subvectors $\left\{\bm{x}_\ell\right\}_{\ell \in \{1,\cdots, L\}}$. A semicolon $(;)$ is used to concatenate column vectors, such that $\left[\bm{a}^\top, \bm{b}^\top\right] = [\bm{a}; \bm{b}]$. Fraktur capital letters denote graphs, defined as $\mathfrak{G} = \left( \mathbb{V}, \mathbb{E} \right)$, where $\mathbb{V}$ and $\mathbb{E}$ are the sets of vertices and edges, respectively.

In this paper, we consider a \ac{mas} composed of $L$ agents, indexed by $\mathbb{L}= \left\{ 1, 2, \ldots, L\right\}$. The discrete-time \ac{lti} dynamics of each agent~$\ell \in \mathbb{L}$ is described by
\begin{align}
    \label{eq:singleagentSS}
    \bm{x}_\ell(t\!+\!1)=\bm{A}_\ell\bm{x}_\ell(t)+\bm{B}_\ell\bm{u}_\ell(t),
\end{align}
where, similar to the notation found in the literature, $t$ denotes the discrete-time index, $\bm{x}_\ell \in \mathbb{R}^n$ is the state vector, $\bm{u}_\ell \in \mathbb{R}^m$ is the control input vector, $\bm{A}_\ell \in \mathbb{R}^{n\times n}$ is the state matrix, and $\bm{B}_\ell \in \mathbb{R}^{n \times m}$ is the control matrix. As is customary in the literature, we assume that the system is stabilizable.

\begin{assumption}
The system in \eqref{eq:singleagentSS} is stabilizable; that is, all eigenvalues of $\bm{A}_\ell$ with modulus greater than or equal to one are controllable.
\end{assumption}

Considering the dynamics of all agents, the \ac{mas} can be expressed in a compact form as
\begin{align}
    \label{eq:masSS}
    \bm{X}(t\!+\!1)=\mathcal{A}\bm{X}(t)+\mathcal{B}\bm{U}(t),
\end{align}
where $\bm{X}=\mathrm{col}\left(\bm{x}_1, \bm{x}_2, \ldots, \bm{x}_{L} \right)$ is the global state vector, and $\bm{U}=\mathrm{col}\left(\bm{u}_1, \bm{u}_2, \ldots, \bm{u}_{
L} \right)$ is the global control input vector. The block matrices $\mathcal{A}\in\mathbb{R}^{nL\times nL}$ and $\mathcal{B}\in\mathbb{R}^{nL\times mL}$ are constructed from $\{\bm{A}_\ell\}_{\ell\in\mathbb{L}}$ and $\{\bm{B}_\ell\}_{\ell\in\mathbb{L}}$ and their interconnections as
\begin{align}
\mathcal{A} &=
\begin{bmatrix}
\bm{A}_1 & \cancel{0} & \cdots & \cancel{0} \\
\cancel{0} & \bm{A}_2 & \cdots & \cancel{0} \\
\vdots & \vdots & \ddots & \vdots \\
\cancel{0} & \cancel{0} & \cdots & \bm{A}_L
\end{bmatrix},
\quad
\mathcal{B} =
\begin{bmatrix}
\bm{B}_1 & \cancel{0} & \cdots & \cancel{0} \\
\cancel{0} & \bm{B}_2 & \cdots & \cancel{0} \\
\vdots & \vdots & \ddots & \vdots \\
\cancel{0} & \cancel{0} & \cdots & \bm{B}_L
\end{bmatrix},
\end{align}
where $\cancel{0}$ denotes a matrix of compatible dimension that may contain nonzero entries. In general, matrices $\mathcal{A}$ and $\mathcal{B}$ may contain nonzero off-block-diagonal terms, due to the inter-agent state couplings or control couplings, as we consider in this paper and discussed in~\cite{shamma2008cooperative,massioni2009distributed,de2006decentralized,ding2019survey}. 
%This paper focuses on such interconnected systems.

To control Agent~$\ell$ in \eqref{eq:singleagentSS}, we consider a quadratic cost function such as
\begin{align}
    \label{eq:quadraticcost}
    g_\ell = \sum_{t=0}^{\infty} \left( \bm{x}_\ell(t)^{\top} \bm{S}_\ell \bm{x}_\ell(t) + \bm{u}_\ell(t)^{\top} \bm{R}_\ell \bm{u}_\ell(t) \right),
\end{align}
where $\bm{S}_\ell \in \mathbb{R}^{n \times n}$ and $\bm{R}_\ell \in \mathbb{R}^{m \times m}$ are positive semi-definite weighting matrices. It is well known that the control policy minimizing the cost function in \eqref{eq:quadraticcost} is a linear state-feedback law~\cite{ogata2002modern}, given by
\begin{align}
    \label{eq:optimalcontorl}
    \bm{u}_\ell(t) = -\bm{K}_\ell \bm{x}_\ell(t),
\end{align}
where $\bm{K}_\ell=\left(\bm{R}_\ell + \bm{B}_\ell^{\top} \bm{P}_\ell \bm{B}_\ell\right)^{-1} \bm{B}_\ell^{\top} \bm{P}_\ell \bm{A}_\ell$ and $\bm{P}_\ell$ is the unique positive definite solution to the discrete-time algebraic Riccati equation
\begin{align}
    \label{eq:riccati}
    &\bm{P}_\ell = \\
    \nonumber&\bm{A}_\ell^{\top} \bm{P}_\ell \bm{A}_\ell - \bm{A}_\ell^{\top} \bm{P}_\ell \bm{B}_\ell \left( \bm{R}_\ell + \bm{B}_\ell^{\top} \bm{P}_\ell \bm{B}_\ell \right)^{-1} \bm{B}_\ell^{\top} \bm{P}_\ell \bm{A}_\ell + \bm{S}_\ell. 
\end{align}

Similarly, to control the \ac{mas} described in \eqref{eq:masSS}, we collect all agents' quadratic cost functions in a compact quadratic cost function 
\begin{align}
    \label{eq:masquadraticcost}
    G= \sum_{\ell=1}^{L} g_\ell = \sum_{t=0}^{\infty} \left( \bm{X}(t)^{\top} \mathcal{S} \bm{X}(t) + \bm{U}(t)^{\top} \mathcal{R} \bm{U}(t) \right).
\end{align}
Here, $\mathcal{S}$ and $\mathcal{R}$ are block-diagonal matrices constructed from $\left\{\bm{S}_\ell\right\}_{\ell \in \mathbb{L}}$ and $\left\{\bm{R}_\ell\right\}_{\ell \in \mathbb{L}}$, respectively. Similarly, the linear feedback control law 
\begin{align}
    \label{eq:masU}
    \bm{U}(t) = -\mathcal{K} \bm{X}(t),
\end{align}
minimizes the cost in \eqref{eq:masquadraticcost} for the \ac{mas} in \eqref{eq:masSS}. The control law in \eqref{eq:masU} stacks the control inputs of all agents. Accordingly, the control input of Agent~$\ell$ is given by
\begin{align}
    \label{eq:agentu}
    \bm{u}_\ell(t) = -\bm{K}_\ell \bm{X}(t),
\end{align}
where $\bm{K}_\ell$ denotes the $\ell^\mathrm{th}$ row block of $\mathcal{K}$. However, $\mathcal{K}$ cannot be computed analytically since $\mathcal{A}$ and $\mathcal{B}$ are unknown in the presence of inter-agent state or control couplings.

As shown in \eqref{eq:agentu}, each agent requires access to the global state vector $\bm{X}$ to execute its control policy. To enable such access, the \ac{mas} connects its agents through a communication network that allows data exchange among them. The network is modeled as an undirected graph $\mathfrak{G} = \left( \mathbb{L}, \mathbb{E}\right)$, where $\mathbb{L}$ is the set of agents (graph nodes) and $\mathbb{E}$ represents the set of edges. An edge $e_{\ell,m} \in \mathbb{E}$ exists between Agents~$\ell$ and~$m$ if a communication link is established between them.

\begin{figure}[t]
\centering
\includegraphics[width=0.9\linewidth]{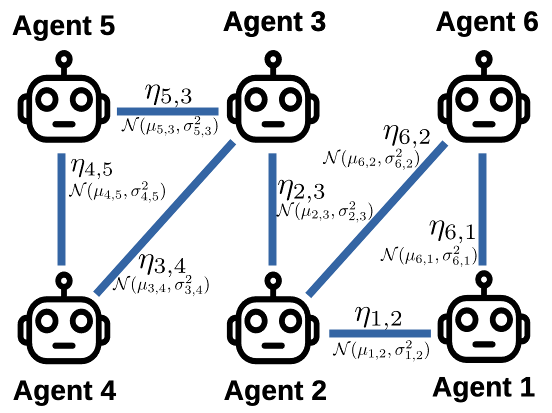}
\caption{Example of a communication network graph}
\label{fig:fig1}
\end{figure}

\begin{assumption}
\label{asmpt:Gconnectivity}
The communication network of the \ac{mas}, represented by the graph $\mathfrak{G}$, is assumed to be connected at all times.
\end{assumption}

\begin{assumption}
\label{asmpt:Gstatic}
The network graph $\mathfrak{G}$ is time invariant, i.e., the network configuration is static and the set of edges remains unchanged.
\end{assumption}
Assumptions~\ref{asmpt:Gconnectivity} and~\ref{asmpt:Gstatic} imply that the network is static and that a communication route exists between any pair of agents. Consequently, each agent can propagate its data to all other agents through multi-hop communication.
%\footnote{A \emph{hop} refers to a single direct communication link between two connected agents, and data transmitted across multiple intermediate agents is said to traverse multiple hops.}.
Specifically, each agent broadcasts its own data to neighbors and relays information received from other agents.

\begin{assumption}
\label{asmpt:Gnocycles}
All routes between any two agents are simple paths; i.e., no route contains repeated nodes.
\end{assumption}
Assumption~\ref{asmpt:Gnocycles} eliminates cycles and ensures that the set of routes is finite. We define the route between Agents~$\ell$ and~$m$ as a set of links, denoted by $\mathbb{R}_{\ell,m}$. The number of links (hops) in a route represents its length and is denoted by $R_{\ell,m}$. The $k^{\mathrm{th}}$ link in the route, denoted by $\mathbb{R}_{\ell,m}^{k}$, corresponds to the edge $\epsilon_{i,j} \in \mathbb{E}$. Any agent~$m$ that has the shortest route to Agent~$\ell$ consisting of a single hop, i.e., $R_{\ell,m}=1$, is considered a neighbor of Agent~$\ell$. The set of neighbors of Agent~$\ell$ is defined as
\begin{align}
    \label{eq:neighbor}
    \mathbb{I}_\ell = \left\{m | m \in \mathbb{L},  R_{\ell,m}=1\right\}.
\end{align}
%All routes in the network are assumed to be known either at design time or determined by routing algorithms such as~\cite{gbadamosi2020design,septiana2016evaluation}.

Additionally, data collection is subject to uncertainty arising from two sources: (\emph{i}) Gaussian channel noise with zero mean, and (\emph{ii}) sensing uncertainty, which can be modeled as a Gaussian process with a non-zero mean~\cite{ISO_GUM_2008}. The sum of these two components results in a Gaussian noise with a non-zero mean. Accordingly, the data collection uncertainly on each link is conservatively modeled as a \ac{nawgn}~\cite{shomorony2013worst}, which captures both stochastic channel noise and sensing bias introduced by sensor imperfections. The non-zero mean value, typically caused by sensor calibration errors, is known from the sensor specifications~\cite{ISO_GUM_2008}.
%we model the noise on each link as \ac{nawgn}, which serves as a conservative representation of the worst-case communication uncertainty~\cite{shomorony2013worst}. 
Note that the noises on different links are independent. Let $\eta_{\ell,m}$ denote the noise associated with the link $e_{\ell,m}$ connecting Agents~$\ell$ and~$m$. 
%Since $e_{\ell,m}$ and $e_{m,\ell}$ represent the same bidirectional link, the corresponding noise terms are equivalent, i.e., $\epsilon_{\ell,m}=\epsilon_{m,\ell}$.
The noise on each link is modeled as
\begin{align}
    \label{eq:Gaussiannoise}
    \eta_{\ell,m} \sim \mathcal{N}(\mu_{\ell,m}, \sigma^2_{\ell,m}),
\end{align}
where $\mu_{\ell,m}$ and $\sigma^2_{\ell,m}$ denote the mean and variance of the noise, respectively.

Fig.~\ref{fig:fig1} shows an example \ac{mas} network graph. This graph is static and connected; for example, a possible route between Agents~$2$ and $4$ is $\mathbb{R}_{2,4} = \left\{ e_{2,3}, e_{3,5}, e_{5,4}\right\}$, consisting of three hops, i.e., $R_{2,4}=3$. The neighbors of Agent~$2$ are $\mathbb{I}_2 = \{1,3,6\}$. The first, second, and third links in the route correspond to 
$\mathbb{R}_{2,4}^1 = e_{2,3}$, $\mathbb{R}_{2,4}^2 = e_{3,5}$, and $\mathbb{R}_{2,4}^3 = e_{5,4}$, respectively, with their associated noise terms denoted by $\eta_{2,3}$, $\eta_{3,5}$, and $\eta_{5,4}$.

\section{Message Passing}
\label{sec:messagepassing}
Control in the \ac{mas} is enabled by data exchange over the communication network, where each agent independently estimates the global state. This is achieved by simulating sampled observations in a distributed manner using the message-passing mechanism. The mechanism first determines the communication routes in the network and then filters the link noise. Finally, it simulates the sampling process to estimate and shift the relative clocks of the agents to compensate communication asynchronous sampling and delays during the learning correction.

\begin{algorithm}[t]
\DontPrintSemicolon
\caption{Noise-Aware Routing within Message-passing Mechanism}
\label{alg:routing}

\KwIn{Graph $\mathfrak{G} = (\mathbb{L}, \mathbb{E})$; link noise parameter $\sigma^2_{i,j}$ for all $e_{i,j} \in \mathbb{E}$; weighting coefficient $\lambda$}
\KwOut{Static routing sets $\mathbb{R}_{\ell,m}$ for all agent pairs $(\ell, m)$}

\ForEach{agent $\ell \in \mathbb{L}$}{
    Initialize $\mathbb{R}_{\ell, m} \gets \emptyset$ $\forall m \in \mathbb{L}$\;
    Initialize $J(\mathbb{R}_{\ell, m}) \gets \infty$ $\forall m \in \mathbb{L}$\;
    Set $J(\mathbb{R}_{\ell,\ell}) \gets 0$\;
    Initialize priority queue $\mathbb{Q} \gets \{\ell\}$\;

    \While{$\mathbb{Q}$ not empty}{
        Extract node $i \gets \arg\min_{v \in \mathcal{Q}} J(\mathbb{R}_{\ell,v})$\;
        Remove $i$ from $\mathbb{Q}$\;
        \ForEach{neighbor $j$ of $i$}{
            Compute link cost:
            \[
                \tilde{\mathbb{R}}_{\ell,j} \gets \mathbb{R}_{\ell,i} \cup \{e_{i,j}\}
            \]
            \[
                J(\tilde{\mathbb{R}}_{\ell,j}) \gets J(\mathbb{R}_{\ell,i}) + (1 + \lambda \sigma^2_{i,j}) \label{line:linkcost}
            \]
            \If{$J(\tilde{\mathbb{R}}_{\ell,j}) < J(\mathbb{R}_{\ell,j})$}{
                Update  $\mathbb{R}_{\ell,j} \gets \tilde{\mathbb{R}}_{\ell,j}$\;
                Update $J(\mathbb{R}_{\ell,j}) \gets J(\tilde{\mathbb{R}}_{\ell,j})$\;
                Add $j$ to $\mathbb{Q}$\;
            }
        }
    }
}
\Return{$\mathbb{R}_{\ell,m}$}\;
\end{algorithm}

\textbf{Routing:}
Data transmission in the \ac{mas} network follows a static routing scheme determined using a Dijkstra-based algorithm~\cite{gbadamosi2020design}, as shown in Alg.~\ref{alg:routing}.
In this context, the selected route minimizes a routing cost~$J(\mathbb{R})$, as shown in Line~\ref{line:linkcost} of Alg.~\ref{alg:routing}. The cost function can be defined to capture relevant communication constraints. In this work, the routing cost is formulated as a Lagrangian combination of the hop count and the link noise. Consequently, the route between Agents~$\ell$ and~$m$ is defined as
\begin{align}
\label{eq:pathLagrangian}
\mathbb{R}_{\ell,m} = \min_{\mathbb{R}} \underbrace{\sum_{e_{i,j} \in \mathbb{R}} \left( 1 + \lambda \sigma^2_{i,j} \right)}_{J(\mathbb{R})},
\end{align}
where $\lambda$ is a Lagrangian weighting coefficient that balances the effects of link noise and communication delay. Note that the mean~$\mu$ is not a factor in the cost function, since its effect can be removed by bias correction.

\begin{figure}[t]
\includegraphics[width=\linewidth]{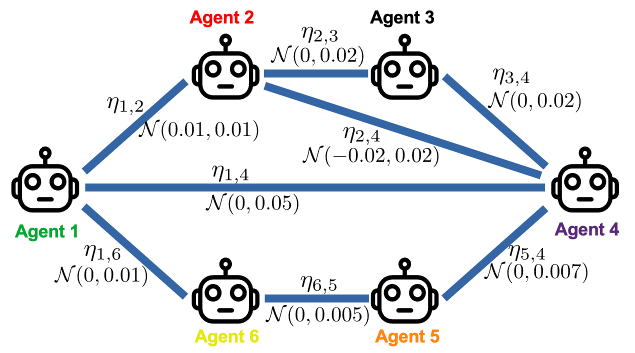}
\caption{Example of route selection in a multi-agent network}
\label{fig:example_messagepassing}
\end{figure}

Fig.~\ref{fig:example_messagepassing} shows an example multi-agent network consisting of six agents. Between Agents~1 and~4, multiple feasible routes exist, including
Route 1: $\{e_{1,4}\}$, Route 2: $\{e_{1,2}, e_{2,4}\}$, Route 3: $\{e_{1,2}, e_{2,3}, e_{3,4}\}$, and Route 4: $\{e_{1,6}, e_{6,5}, e_{5,4}\}$. 
Algorithm~\ref{alg:routing} selects the minimum-cost route according to \eqref{eq:pathLagrangian}. Considering the link noise parameters shown in the figure, the resulting route costs for each of these routes are given by
\begin{align}
\nonumber J_1 &= 1 + 0.05\lambda, \\
\nonumber J_2 &= 2 + 0.03\lambda, \\
\nonumber J_3 &= 3 + 0.05\lambda, \\
J_4 &= 3 + 0.022\lambda,
\end{align}
respectively.
When $\lambda=1$, the corresponding cost values are $1.05$, $2.03$, $3.05$, and $3.022$, respectively, and therefore Route 1 is optimal. %route is $\mathbb{R}_{1,4}=\{e_{1,4}\}$. 
For $\lambda=100$, the corresponding costs are $6$, $5$, $8$, and $5.2$, respectively, making Route~2 the optimal route.
Finally, when $\lambda=500$, the costs become $26$, $17$, $28$, and $14$, respectively, and the optimal route is Route 4.

\textbf{Noise refinement:}
Each agent~$\ell$ at Time $t$ receives the state information from another agent~$m$ via the predetermined route~$\mathbb{R}_{\ell,m}$. The information received from Agent~$m$ is denoted by $\bm{y}_m(t)$ and given by
\begin{align}
    \label{eq:recievedstate}
    \bm{y}_m(t) &= \bm{x}_m(t) + \underbrace{\sum_{e_{i,j}\in \mathbb{R}_{\ell,m}} \eta_{i,j}}_{\eta_{\mathrm{total}}},
\end{align}
where $\bm{x}_m(t)$ is the actual state information. Here, $\eta_{\mathrm{total}} \sim \mathrm{N}(\mu_{\mathrm{total}}, \sigma_{\mathrm{total}}^2)$ is the accumulated noise which is also \ac{nawgn} with the parameters
\begin{align}
    \label{eq:sumofGaussianmedian}
    \mu_{\mathrm{total}} = \sum_{e_{i,j}\in \mathbb{R}_{\ell,m}} \mu_{i,j}, \qquad \sigma_{\mathrm{total}}^2 = \sum_{e_{i,j}\in \mathbb{R}_{\ell,m}} \sigma_{i,j}^2.
\end{align}
With the received state information~$\bm{y}_m(t)$, each agent~$\ell$ constructs its observed global state vector as
\begin{align}
    \bm{Y}_\ell(t) = \mathrm{col}\left( \bm{y}_1(t), \ldots, \bm{x}_\ell(t),\ldots, \bm{y}_L(t)\right).
\end{align}
Note that each agent~$\ell$ accesses its own state information noise-free.

Since the network configuration is static, the message-passing mechanism at each agent refines $\bm{y}_m(t)$ upon reception. First, the accumulated bias~$\mu_{\mathrm{total}}$ is removed via mean adjustment using
\begin{equation}
    \label{eq:mean_adjusted}
    \bar{\bm{y}}_m(t) = \bm{y}_m(t) - \mu_{\mathrm{total}},
\end{equation}
which ensures unbiased observations prior to temporal filtering.

The bias-corrected signal~$\bar{\bm{y}}_m(t)$ is then denoised using a recursive first-order exponential low-pass filter in the form of an \ac{ema},
\begin{equation}
    \label{eq:ema_filter}
    \tilde{\bm{x}}_m(t) = \beta(t)\,\bar{\bm{y}}_m(t) + \big(1 - \beta(t)\big)\,\tilde{\bm{x}}_m(t-1),
\end{equation}
where $\tilde{\bm{x}}(t)$ represents the refined state information at Time $t$, and $\beta(t) \in (0,1]$ is the adaptive smoothing coefficient that governs the trade-off between responsiveness to new information and noise attenuation.

Unlike model-based Bayesian filters, the EMA operates without requiring explicit system dynamics, observation models, or noise covariance matrices. This model-free structure significantly reduces computational and implementation complexity, making the filter particularly attractive for large-scale or resource-constrained multi-agent networks. Moreover, the recursive form enables online operation with constant memory and computation per update.

This filter is particularly well suited for the considered \ac{mas} setting, in which the system states evolve smoothly over time and the aggregated link noise is approximately white and uncorrelated. The \ac{ema} resembles the scalar steady-state Kalman filter update for a random-walk process under simplifying assumptions~\cite{bishop2001introduction}. However, it is not implemented as a Kalman filter, since no explicit system dynamics, observation model, or noise covariance matrices are assumed or propagated. As a result, the proposed approach provides an efficient, model-free denoising solution that avoids covariance estimation and Riccati recursions while retaining the benefits of recursive optimal linear smoothing in steady state.

The adaptive smoothing coefficient $\beta(t)$ in \eqref{eq:ema_filter} is determined from the steady-state Kalman gain expression,
\begin{equation}
    \label{eq:beta_selection}
    \beta(t) = \frac{\sigma_x^2(t)}{\sigma_x^2(t) + \sigma_{\mathrm{total}}^2(t)},
\end{equation}
where $\sigma_x^2(t)$ denotes the empirical variance of the refined state estimate. This variance is updated online using a short rolling window of the most recent $W$ samples:
\begin{align}
    \label{eq:variance_estimation}
        \sigma_x^2(t) & = \mathrm{Var}\big(\tilde{\bm{x}}(t) - \tilde{\bm{x}}(t-1)\big),
\end{align}
where $\mathrm{Var}(\cdot)$ denotes the empirical unbiased estimation of the variance. To maintain numerical stability, $\beta(t)$ can be constrained.

With this filtered state information, each agent~$\ell$ constructs the refined global state estimate as
\begin{align}
    \label{eq:refinedglobalstates}
    \bm{\tilde{X}}(t) = \mathrm{col}\left( \bm{\tilde{x}}_1(t),\ldots, \bm{x}_\ell(t), \ldots, \bm{\tilde{x}}_L(t) \right),
\end{align}
and use it as an immediate observation to learn the local control policy.

Following the example in Fig.~\ref{fig:example_messagepassing}, when $\lambda=100$, the selected route from Agent~1 to Agent~4 is Route 2: $\{e_{1,2}, e_{2,4}\}$. Agent~1 receives
\begin{align}
\bm{y}_4(t)
= \bm{x}_4(t) + \underbrace{\eta_{1,2} + \eta_{2,4}}_{\eta_{\mathrm{total}}},
\end{align}
and, under the independence assumption, the aggregated noise is calculated as $\eta_{\mathrm{total}} \sim \mathcal{N}(-0.01,0.03)$ using \eqref{eq:sumofGaussianmedian}. To illustrate the refinement, consider that the received information is a single state component and at Time $t$, its value is $y_4(t) = 1.98$. The bias-adjusted observation in~\eqref{eq:mean_adjusted} becomes
\begin{equation}
\bar{y}_4(t)
= y_4(t) - \mu_{\mathrm{total}}
= 1.98 - (-0.01)
= 1.99,
\end{equation}
which is unbiased, with residual noise distributed as
$\mathcal{N}(0, 0.03)$.

For denoising, let the previous refined state estimate be $\tilde{x}_4(t-1) = 1.95$ and choose $\beta(t) = 0.2$. Applying the \ac{ema} filter in~\eqref{eq:ema_filter}, then the refined estimate at Time $t$ is 
\begin{align}
\nonumber\tilde{x}_4(t) &= \beta(t)\,\bar{y}_4(t) + \big(1-\beta(t)\big)\tilde{x}_4(t-1)\\
\nonumber&= 0.2(1.99) + 0.8(1.95)\\
&= 1.958.
\end{align}

\begin{algorithm}[t]
\DontPrintSemicolon
%\SetAlgoNoLine
\SetKwInput{KwInput}{Input}
\SetKwInput{KwOutput}{Output}

\KwInput{Relative timings $d_{\ell,m}$ for each agent $m$'s route; window size $P$; refined global state information $\bm{\tilde{X}}$}
\KwOutput{FIFO Buffer $\mathbb{B}_\ell$}

\textbf{Initialization:}
$\mathbb{B}_\ell \gets$ FIFO queue (capacity $P$)\;
%\ForEach{$m \in \mathbb{L}$}{
%  $d_{\ell,m} \gets R_{\ell,m} - 1$\;
%}
\BlankLine

\While{\KwSty{not} done}{

receive $\bm{\tilde{X}}(t)$\;

%\ForEach{item $\bm{\tilde{x}}_i(t)$ in $\bm{\tilde{X}}(t)$}{
\ForEach{$m$ in $\mathbb{L}$}{
    $\bm{Z} \gets$ pick the $d_{\ell,m}$-th element from $\mathbb{B}_\ell$ \label{line:pickdata}\;
    replace $\bm{\hat{x}}_m \in \bm{Z}$ with $\bm{\tilde{x}}_m(t) \in \bm{\tilde{X}}(t)$ \label{line:reconstruct} \label{line:replacedata} \;
    update the $d_{\ell,m}$-th element from $\mathbb{B}_\ell$ with $\bm{Z}$\;
}
pop the oldest element from $\mathbb{B}_\ell$\;
push back $\bm{Z} = 0_L$ into $\mathbb{B}_\ell$ \label{line:buffer} \; 
}
\Return{$\mathbb{B}$}\;

\caption{Time-shifting of the refined global state estimate at each agent}
\label{alg:async-mp}
\end{algorithm}

\textbf{Relative time shift:}
State information in the described \ac{mas} is observed and exchanged in the discrete-time domain through local sampling at each agent. In the absence of a shared global clock, agents do not sample their states synchronously; instead, each agent operates based on its own local time reference. As a result, the sampled state information exchanged over the network is inherently asynchronous, and all transmitted data are sampled with respect to the sender’s local clock.

In addition to asynchronous sampling, communication delays further degrade the temporal consistency of shared information. In particular, agents not only receive the state information with offset timestamps, but also receive it after a delay. Since the network configuration and routings are static, the state information exchanged between agents is transmitted over fixed multi-hop routes, with each hop introducing a deterministic delay. Consequently, for any pair of agents, a \emph{relative clock} can be defined that captures both the cumulative communication delay and the offset between local timestamps.

Specifically, regardless of the sender’s local timestamp or the absolute reception time at the receiving agent, the hop count of the routing uniquely determines the number of discrete time steps in the past at which the transmitted state information was sampled. Consequently, each received state can be unambiguously associated with a specific time step offset with respect to the current local time of the receiving agent. Rather than enforcing explicit clock synchronization, the proposed message-passing mechanism performs \emph{relative time alignment} by compensating for hop-dependent delays. This approach avoids clock-estimation protocols and relies solely on routing information.

With each forwarding node introducing a single discrete-time delay step and $R_{\ell,m}$ denoting the hop counts along the route from Agent~$m$ to Agent~$\ell$, the corresponding relative timing offset is defined as $d_{\ell,m} = R_{\ell,m} - 1$. As a result, direct communication links incur zero relative timing offset, while longer routes introduce proportionally larger offsets. Since routing is static, these offsets remain constant throughout the operation.

The message-passing mechanism exploits these relative timing offsets through a time-shifting process, as described in Alg.~\ref{alg:async-mp}. Specifically, each agent~$\ell$ maintains a FIFO buffer $\mathbb{B}_\ell$ of capacity~$P$, which stores time-shifted global state estimates. Upon receiving the refined global state estimate~$\bm{\tilde{X}}(t)$, the agent performs the following steps:
\begin{enumerate}
    \item Each component of the refined global state estimate, i.e., ~$\bm{\tilde{x}}_m(t)$, is assigned to the buffer entry corresponding to its hop-induced delay $d_{\ell,m}$ as shown in Line~\ref{line:pickdata} of Alg.~\ref{alg:async-mp};
    \item The component~$\bm{\tilde{x}}_m(t)$ replaces the corresponding component in the refined global state estimate at that buffer index~$\bm{\hat{z}}_m(t)$ as shown in Line~\ref{line:replacedata} of Alg.~\ref{alg:async-mp}.
\end{enumerate}
This procedure effectively shifts each agent’s state estimate backward in time, aligning all components to a common temporal time reference without requiring clock synchronization. The time-shifted global state estimates are updated recursively and stored in the buffer as new information becomes available.

\begin{table}[!t]
\caption{Example of routing in the MAS network of Fig.~\ref{fig:example_messagepassing}}
\label{tab:exampleroutes}
\centering
\begin{tabular}{|c||c|c|c|}
\hline
Receiver & Sender & Route & Hop count\\
\hline
Agent~1& Agent~2 & $\{e_{1,2}\}$ & 1\\ \hline
Agent~1& Agent~3 & $\{e_{1,2}, e_{2,3}\}$ & 2\\ \hline
Agent~1& Agent~4 & $\{e_{1,2}, e_{2,4}\}$ & 2\\ \hline
Agent~1& Agent~5 & $\{e_{1,4}, e_{4,5}\}$ & 2\\ \hline
Agent~1& Agent~6 & $\{e_{1,6}\}$ & 1\\
\hline
\end{tabular}
\end{table}

Consider again the \ac{mas} network shown in Fig.~\ref{fig:example_messagepassing}. When $\lambda=100$, Agent~1 receives state information from other agents through the routes shown in Table~\ref{tab:exampleroutes}.
%\begin{align}
%    \nonumber&\mathbb{R}_{1,2} = \{e_{1,2}\}\quad
%    \nonumber\mathbb{R}_{1,3} = \{e_{1,2}, e_{2,3}\}\quad
%    \nonumber\mathbb{R}_{1,4} = \{e_{1,2}, e_{2,4}\}\\
%    &\mathbb{R}_{1,5} = \{e_{1,4}, e_{4,5}\}\quad
%    \mathbb{R}_{1,6} = \{e_{1,6}\}
%\end{align}
%which correspond to hop counts $R_{1,2}=1$, $R_{1,3}=2$, $R_{1,4}=2$, $R_{1,5}=2$, and $R_{1,6}=1$, respectively.
At Time $t$, Agent~1 receives the refined global state estimate from other agents, computed according to \eqref{eq:refinedglobalstates} as
\begin{align}
    \bm{\tilde{X}}(t) = \mathrm{col}\left( 0.12,\, 0.142,\, 0.633,\, 1.958,\, 0.32,\, 1.531 \right).
\end{align}
However, this information is both delayed and sampled asynchronously; consequently, individual state components may correspond to different past discrete time steps. Since hop counts are known, the relative timing for each state component is known and is denoted by $d_{\ell,m}$.

In this example, since $d_{1,2}=d_{1,6}=0$, the refined state estimate components of Agents~1, 2, and 6, namely~$\bm{\tilde{x}}_1(t)$, $\bm{\tilde{x}}_2(t)$, and $\bm{\tilde{x}}_6(t)$, correspond to the same time step~$t$. In contrast, the components of Agents~3, 4, and 5, namely~$\bm{\tilde{x}}_3(t)$, $\bm{\tilde{x}}_4(t)$, and $\bm{\tilde{x}}_5(t)$, correspond to $t-1$. As a result, the components~$\bm{\tilde{x}}_3(t)$, $\bm{\tilde{x}}_4(t)$, and $\bm{\tilde{x}}_5(t)$ are shifted backward by one time step, while the information received subsequently at $t+1$ is aligned with the current time reference~$t$.

Using Alg.~\ref{alg:async-mp}, Agent~1 maintains buffer $\mathbb{B}_1$ and applies the shifting process to all components of the received refined state estimate. This process is repeated recursively as new information arrives, enabling the agent to maintain a buffer of time-shifted refined global state estimates for learning correction.
The contents of buffer~$\mathbb{B}_1$ prior to applying the shifting process are given by
\begin{align}
    \nonumber &\vdots\\
    \nonumber\bm{\tilde{X}}(t+1) &= \mathrm{col}\left( 0.115,\, 0.128,\, 0.601,\, 1.912,\, 0.309,\, 1.531 \right),\\
    \nonumber\bm{\tilde{X}}(t) &= \mathrm{col}\left( 0.120,\, 0.142,\, 0.633,\, 1.958,\, 0.320,\, 1.386 \right),\\
    \nonumber\bm{\tilde{X}}(t-1) &= \mathrm{col}\left( 0.118,\, 0.166,\, 0.694,\, 1.893,\, 0.388,\, 1.247 \right),\\
    &\vdots.
\end{align}
After recursively applying the time-shifting process, the buffer is updated as
\begin{align}
    \nonumber &\vdots\\
    \nonumber\bm{\tilde{X}}(t+1) &= \mathrm{col}\left( 0.115,\, 0.128,\, 0.588,\, 2.011,\, 0.377,\, 1.531 \right),\\
    \nonumber\bm{\tilde{X}}(t) &= \mathrm{col}\left( 0.120,\, 0.142,\, 0.601,\, 1.912,\, 0.309,\, 1.386 \right),\\
    \nonumber\bm{\tilde{X}}(t-1) &= \mathrm{col}\left( 0.118,\, 0.166,\, 0.633,\, 1.958,\, 0.320,\, 1.247 \right),\\
    &\vdots.
\end{align}

\begin{algorithm}[t]
\DontPrintSemicolon
\caption{Encoded corrective double deep Q-network (CDNet)}
\label{alg:rl}
\KwIn{Agents $\mathbb{L}$; }
\KwOut{Control policy~$\mathcal{K}$}

\textbf{Initialization:}\;
run Alg.~\ref{alg:routing}\;
init trunk (global encoder) $\bm{\Phi}(\boldsymbol{\tilde{X}}) \longmapsto \bm{\phi}$\;\label{line:encoder}
\ForEach{$\ell \in \mathbb{L}$}{
  init encoder head $\bm{\Psi}_\ell(\boldsymbol{\phi}) \longmapsto \bm{\psi}_\ell$\;
  init actor  $\boldsymbol{\kappa}_\ell(\boldsymbol{\psi}_\ell)$\;
  init critics $Q_\ell(\boldsymbol{\psi}_\ell, \bm{K}_\ell | \theta_\ell)$ and $Q_\ell^\prime(\boldsymbol{\psi}_\ell, \bm{K}_\ell | \theta_\ell^\prime)$\;
  init buffer $\mathbb{H}_\ell$\;
}
\BlankLine
\For{each episode}{
    Reset environment\;
    run Alg.~\ref{alg:async-mp} in parallel\;
    \While{\KwSty{not} done}{
            \ForEach{agent $\ell$}{
                Generate action: $\bm{K}_\ell^\ast \gets$ sample $\boldsymbol{\kappa}_\ell$\; 
            }
            calculate and apply $\bm{U}$\;
            update $\bm{\Phi}(\bm{\tilde{X}})$ with $\bm{\tilde{X}}_\ell$ from all agents\;
            \ForEach{agent $\ell$}{
                calculate $r_\ell$, $\hat{Q}_\ell$, and $\rho_\ell$\;\label{line:rhor}
                update $Q_\ell$ and $Q^\prime_\ell$ by minimizing~$L_\ell$ and $L_\ell^\prime$\;\label{line:updatecritics}
                update $\bm{\kappa}_\ell$ to maximize $\hat{Q}_\ell$\label{line:updatek}
            }
            \ForEach{agent $\ell$}{
                \If{$D_\ell \times P_\ell$ steps passed}{\label{line:bufferready}
                    \While{replay $\mathbb{B}_\ell$}{
                    $\bm{\hat{X}}_\ell \gets$ pick and pop from $\mathbb{B}_\ell$\;
                    Compute reward correction $r^\prime_\ell$ with $\bm{\hat{X}}_\ell$\;\label{line:samplebuffer}
                    Add $(\bm{\hat{X}}_\ell, r'_\ell)$ to $\mathbb{B}_\ell$\;\label{line:history}
                    }
                }
                \If{$\mathbb{H}_\ell$ not empty}{
                \While{replay $\mathbb{H}_\ell$}{
                Soft update $\boldsymbol{\kappa}_\ell$, $Q_\ell$, and $Q^\prime_\ell$ based on  $r'_\ell$\;\label{line:softupdate}
                }
                }
            }
    }
} 

\ForEach{agent $\ell$}{
    $\bm{K}_\ell \gets$ sample $\boldsymbol{\kappa}_\ell$\;
}
\Return $\mathcal{K}=[\bm{K}_1;\ldots,\bm{K}_L]$\;
\end{algorithm}

\section{Learning Framework}
\label{sec:proposed}
The proposed learning framework is built on a corrective double Q-network approach, as shown in Alg.~\ref{alg:rl}. The framework follows a distributed actor-critic structure in which each agent maintains its own policy (actor) and value (critic) estimators, while a globally shared encoder extracts coupling-aware features form the global state estimates.

A shared encoder trunk~$\boldsymbol{\Phi}(\bm{\tilde{X}})$ that collects all agents' refined global state estimates and maps them to a latent representation~$\boldsymbol{\phi}$, as shown in Line~\ref{line:encoder} of Alg.~\ref{alg:rl}. Note that the refined estimates~$\bm{\tilde{X}}_\ell$ are provided by the message-passing mechanism, as discussed in Section~\ref{sec:messagepassing}.
The purpose of the shared trunk is to extract structured features that capture the coupling among agents. Instead of directly using observations, the learning process operates on this encoded latent representation, which improves robustness.

Each agent~$\ell$ employs an individual encoder head~$\boldsymbol{\Psi}_\ell(\boldsymbol{\phi})$ that produces an agent-specific feature~$\boldsymbol{\psi}_\ell$. This modular structure avoids constructing a single large centralized critic while still preserving global information through the shared trunk. With~$\boldsymbol{\psi}_\ell$, each agent maintains an actor network~$\bm{\kappa}_\ell(\boldsymbol{\psi}_\ell)$ that generates a control gain~$\bm{K}_\ell$. The control input is computed using the locally available refined global state estimate as $\bm{U}_\ell(t) = -\bm{K}_\ell \bm{\tilde{X}}_\ell(t)$.

Furthermore, each agent maintains two critic Q-networks $Q_\ell(\boldsymbol{\psi}_\ell, \bm{K}_\ell | \theta_\ell)$ and $Q_\ell^\prime(\boldsymbol{\psi}_\ell, \bm{K}_\ell | \theta_\ell^\prime)$. These critics approximate the action-value function defined as the expected discounted return, 
\begin{align}
    \label{eq:Qapprox}
    Q_\ell(\boldsymbol{\psi}_\ell, \bm{K}_\ell) = \mathbb{E} \left[ \sum_{t=0}^{\infty} \gamma^{t} \, r_\ell(t) \,\Big|\, \bm{\psi}_\ell(0), \, \bm{K}_\ell (0) \right],
\end{align}
where $\gamma \in (0,1)$ is the discount factor and $r_\ell(t)$ is the immediate reward of Agent~$\ell$, defined as 
\begin{align}
    \label{eq:immiditaereward}
    r_\ell(t) = -\bm{\tilde{X}}_\ell(t)^\top \mathcal{S}_\ell \bm{\tilde{X}}_\ell(t) - \bm{U}_\ell(t)^\top \mathcal{R}_\ell \bm{U}_\ell(t).
\end{align}
These approximate critic values correspond to the quadratic cost in \eqref{eq:quadraticcost} and preserve the structure of the Markov decision process. 

The target value for critic updates is computed as
\begin{align}
    \label{eq:Qvalue}
    \rho_\ell = r_\ell(t) + \gamma \hat{Q}_\ell(\boldsymbol{\psi}_\ell^+, \bm{K}_\ell^\ast),\qquad \bm{K}_\ell^\ast = \boldsymbol{\kappa}_\ell(\boldsymbol{\psi}_\ell^+),
\end{align}
where $\boldsymbol{\psi}_\ell^+$ denotes the next step feature vector and $\hat{Q}_\ell$ is chosen by
\begin{align}
    \label{eq:pessimisticQ}
    \hat{Q}_\ell(\boldsymbol{\psi}_\ell, \bm{K}_\ell) = \min \big( Q_\ell(\boldsymbol{\psi}_\ell, \bm{K}_\ell), Q_\ell^\prime(\boldsymbol{\psi}_\ell, \bm{K}_\ell)\big),
\end{align}
to be the pessimistic reward prediction to mitigate overestimation which is typical in Q-networks when observation is noisy~\cite{overestimate}. This is shown in Line~\ref{line:rhor} of Alg.~\ref{alg:rl}.

The critic networks are updated in Line~\ref{line:updatecritics} of Alg.~\ref{alg:rl} by minimizing the temporal difference Loss $L_\ell(\theta_\ell)$ and $L_\ell^\prime(\theta_\ell^\prime)$, calculated as 
\begin{align}
    \label{eq:qnetworkupdate}
    \nonumber L_\ell(\theta_\ell) = \big( Q_\ell(\boldsymbol{\psi}_\ell, \bm{K}_\ell) - \rho_\ell \big)^2,\\
    L_\ell^\prime(\theta_\ell^\prime) = \big( Q_\ell^\prime(\boldsymbol{\psi}_\ell, \bm{K}_\ell) - \rho_\ell \big)^2.
\end{align}
Simultaneously, the actor~$\bm{\kappa}_\ell$ is updated in the direction of maximizing $\hat{Q}$
to encourage the actor to select gain vectors~$\bm{K}_\ell^\ast$ that lead to high, yet conservative long-term value estimates, as shown in Line~\ref{line:updatek} of Alg.~\ref{alg:rl}. This structure stabilizes learning by reducing value overestimation and improving robustness to noise.

\begin{figure*}[!t]
\begin{align}
\label{eq:Aparam}
\nonumber &\mathcal{A}_{5} =
\left[
\begin{array}{rrrrr}
 0.63 & 0.05 & -0.04 & 0.03 & 0.02\\
 0.04 & 0.59 & 0.06 & -0.03 & 0.05\\
 -0.02 & 0.03 & 0.56 & 0.04 & -0.02\\
 0.03 & -0.02 & 0.02 & 0.58 & 0.06\\
 0.02 & 0.03 & -0.02 & 0.04 & 0.57
\end{array}
\right],\quad
\mathcal{A}_{6} =
\left[
\begin{array}{rrrrrr}
 0.63 & 0.05 & -0.04 & 0.03 & 0.02& -0.01\\
 0.04 & 0.59 & 0.06 & -0.03 & 0.05& 0.00\\
 -0.02 & 0.03 & 0.56 & 0.04 & -0.02& 0.02\\
 0.03 & -0.02 & 0.02 & 0.58 & 0.06& 0.01\\
 0.02 & 0.03 & -0.02 & 0.04 & 0.57& 0.01\\
 0.02 & 0.00 & -0.03 & -0.02 & -0.02 & 0.60
\end{array}
\right],\\
&\mathcal{A}_8 =
\left[
\begin{array}{rrrrrrrr}
 0.63 & 0.05 & -0.04 & 0.03 & 0.02 & -0.01 & 0.02 & -0.01 \\
 0.04 & 0.59 & 0.06 & -0.03 & 0.05 & 0.00 & 0.01 & 0.00 \\
 -0.02 & 0.03 & 0.56 & 0.04 & -0.02 & 0.02 & -0.02 & 0.01 \\
 0.03 & -0.02 & 0.02 & 0.58 & 0.06 & 0.01 & -0.00 & -0.01 \\
 0.02 & 0.03 & -0.02 & 0.04 & 0.57 & 0.01 & 0.00 & 0.01 \\
 0.02 & 0.00 & -0.03 & -0.02 & -0.02 & 0.60 & -0.01 & 0.03 \\
 -0.02 & -0.03 & 0.02 & -0.00 & 0.01 & 0.02 & 0.58 & -0.03 \\
 0.01 & -0.01 & 0.03 & 0.02 & 0.01 & -0.00 & -0.03 & 0.59
\end{array}
\right]
\end{align}
\end{figure*}

While the above procedure addresses noisy observations, asynchronous communication and multi-hop delays have not been considered yet. To compensate for these effects, Alg.~\ref{alg:async-mp} runs in parallel as part of the message-passing mechanism. Each agent maintains a buffer $\mathbb{B}_\ell$ containing time-shifted refined global state estimates~$\bm{\tilde{X}}_\ell(t)$, as discussed in Section~\ref{sec:messagepassing}. Let
\begin{align}
    D_\ell = \max_{m \in \mathbb{L}} d_{\ell,m}
\end{align}
denote the maximum time-shift between Agent~$\ell$ and any other agent. Since each component of the global state requires at most $D_\ell$ steps to align, the buffer becomes fully updated after $D_\ell\times P_\ell$ time steps. Once this condition is satisfied, as shown in Line~\ref{line:bufferready} of Alg.~\ref{alg:rl}, the buffer is replayed and the associated rewards are recomputed using the time-shifted state estimates in \eqref{eq:immiditaereward}, as shown in Line~\ref{line:samplebuffer} of Alg.~\ref{alg:rl}. The recalculated reward, denoted by~$r_\ell^\prime$ represents a reward correction that compensates for induced temporal inconsistencies.

The corrected pairs~$\left(\bm{\hat{X}}_\ell(t),r_\ell^\prime\right)$ are stored in a secondary history buffer~$\mathbb{H}_\ell$, as shown in Line~\ref{line:history} of Alg.~\ref{alg:rl}. Once the buffer~$\mathbb{H}_\ell$ is fully populated, it is replayed to recalculate~$\rho_\ell$ using \eqref{eq:Qvalue}, update both critic networks~$Q_\ell$ and $Q_\ell^\prime$ using \eqref{eq:qnetworkupdate}, and also update the actor~$\boldsymbol{\kappa}_\ell$ accordingly. Unlike the primary online updates, this corrective phase employs soft parameter updates with a reduced learning rate (Line 26) to ensure stability and prevent abrupt parameter shifts, as shown in Line~\ref{line:softupdate} of Alg.~\ref{alg:rl}.

\section{Evaluation}
\label{sec:evaluation}
In this section, we first describe the simulation setup used to evaluate the proposed \ac{coco} algorithm. We then present performance results across multiple test cases and analyze \ac{coco}'s regret, robustness, and scalability. Finally, we compare the performance of \ac{coco} with the related work.

\subsection{Evaluation Setup}
Our proposed \ac{coco} algorithm is implemented in Python and trained on a workstation running Ubuntu with an Intel Xeon Gold 6248 CPU, 128~GB RAM, and two NVIDIA RTX A4000 GPUs. The implementation consists of more than 500 lines and is publicly available on Github\footnote{\ac{coco} is available at Github via \url{https://github.com/kemenghan/Noisy_Q_Correction}.}.

\begin{figure}[t]
  \centering
  \begin{subfigure}{0.23\linewidth}
    \includegraphics[width=\linewidth]{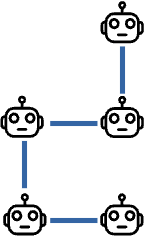}
    \caption{Line}
    \label{fig:line}
  \end{subfigure}
  \hfill
  \begin{subfigure}{0.23\linewidth}
    \includegraphics[width=\linewidth]{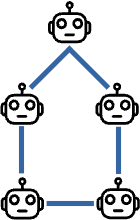}
    \caption{Ring}
    \label{fig:ring}
  \end{subfigure}
  \hfill
  \begin{subfigure}{0.3\linewidth}
    \includegraphics[width=\linewidth]{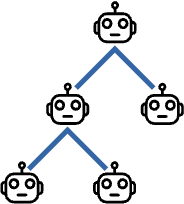}
    \caption{Tree}
    \label{fig:tree}
  \end{subfigure}
  \caption{Examples of different topologies with 5 agents}
  \label{fig:fig4}
\end{figure}

We evaluate \ac{coco} on \acp{mas} with $L \in \{5, 6, 8\}$ agents under three communication topologies: line, ring, and tree. These example topologies are shown in Fig.~\ref{fig:fig4}. 
For simplicity and without loss of generality, each agent is modeled with a scalar state and a scalar control input. Consequently, the dimension of the global state is equal to the number of agents $L$. The system dynamics defined in~\eqref{eq:masSS} uses the state matrices $\mathcal{A}$ given in~\eqref{eq:Aparam}, while the input matrices are chosen as identity matrices~$\mathcal{B} = I_L$. This representation is commonly adopted in the \ac{mas} literature since the global system can be written in the stacked form of \eqref{eq:masSS}. For agents with vector-valued states or inputs, the same formulation holds by interpreting $\mathcal{A}$ and $\mathcal{B}$ as block matrices that capture the coupling between agents. Therefore, modeling agents with scalar states simplifies the exposition while preserving the general structure of the multi-agent dynamics widely used in related work~\cite{li2015fully,aysal2008distributed,zhang2019distributed}.
%These matrices in \eqref{eq:masSS} already capture the general linear interconnection structure. In particular, the matrix $\mathcal{A} \in \mathbb{R}^{nL\times nL}$ can equivalently represent different agent state dimensions through block partitions. For example, a $6 \times 6$ matrix can correspond to six agents with scalar states ($1 \times 1$ blocks), three agents with two-dimensional states ($2 \times 2$ blocks), or two agents with three-dimensional states ($3 \times 3$ blocks). The scalar-agent formulation therefore simplifies the notation while preserving the general structure of the multi-agent dynamics. Similarly, choosing $\mathcal{B} = I_L$ corresponds to each agent having an independent control channel. For higher-dimensional agent models, $\mathcal{B}$ would simply become block diagonal with appropriate block sizes, and the analysis remains unchanged.
For the quadratic cost function defined in~\eqref{eq:masquadraticcost} and the corresponding reward definition in~\eqref{eq:immiditaereward}, we set the state and control weight matrices to identity matrices, $\mathcal{R} =I_L$ and $\mathcal{S} =I_L$. 

The shared encoder consists of two fully connected layers with 64 ReLU-activated units, each mapping the $L$-dimensional refined global state estimation to a 64-dimensional latent representation. Each agent maintains:
\begin{enumerate}
    \item \textbf{One Policy head:} one fully connected hidden layer with 64 ReLU-activated units, followed by a Tanh-activated output layer generating a scalar control action.
    \item \textbf{Two critic heads:} each composed of a 64-unit ReLU-activated layer followed by a linear output layer producing a scalar Q-value for each state–action pair.
\end{enumerate}
All networks are trained with a learning rate of $1\times 10^{-4}$. The learning is conducted for 5,000 episodes, with 10 simulation steps per episode. An experience replay buffer with a capacity of 1,000 transitions is employed, and mini-batches of size 32 are sampled during learning. To examine robustness under different communication conditions, various delay–noise balancing weights $\lambda \in \{1,100,500\}$ are used.

For each combination of agent number and network topology, \ac{nawgn} is introduced on every link. The noise parameters are initialized independently per link, where both mean $\mu_{i,j}$ and variance $\sigma^2_{i,j}$ are independently sampled from a uniform distribution over $[0, 0.1]$. Since the communication graph is undirected, identical noise parameters are assigned to both directions of the same link. As an illustrative example, for the $5$-agent test case under a line topology, the sampled noise distributions are:
\begin{align}
\nonumber\eta_{1,2} = \eta_{2,1} &\sim \mathcal{N}(0.05, 0.02)\\
\nonumber\eta_{2,3} = \eta_{3,2} &\sim \mathcal{N}(0.08, 0.09)\\
\nonumber\eta_{3,4} = \eta_{4,3} &\sim \mathcal{N}(0.03, 0.04)\\
\eta_{4,5} = \eta_{5,4} &\sim \mathcal{N}(0.04, 0.05).
\end{align}
Note that the noise parameters differ across topologies because both the number of communication links and the network configuration vary. For example, in the $5$-agent test case, the line topology contains four links, whereas the ring topology contains five links. Consequently, each topology is associated with a different set of link noise distributions.

\subsection{Simulation Results}
This section presents the simulation results of \ac{coco} algorithm. The objective is to examine (\emph{i}) the stability and convergence behavior of the proposed learning method, (\emph{ii}) the performance distribution among agents, and (\emph{iii}) \ac{mas} control performance. The closed-loop control behavior of \ac{coco} under the learned controller is evaluated. Specifically, performance is quantified using the cumulative quadratic cost defined in \eqref{eq:masquadraticcost}, which directly reflects the quadratic cost of \ac{mas}. This cost is computed for every learning episode.

\begin{figure}[!t]
\centering
\includegraphics[width=\linewidth]{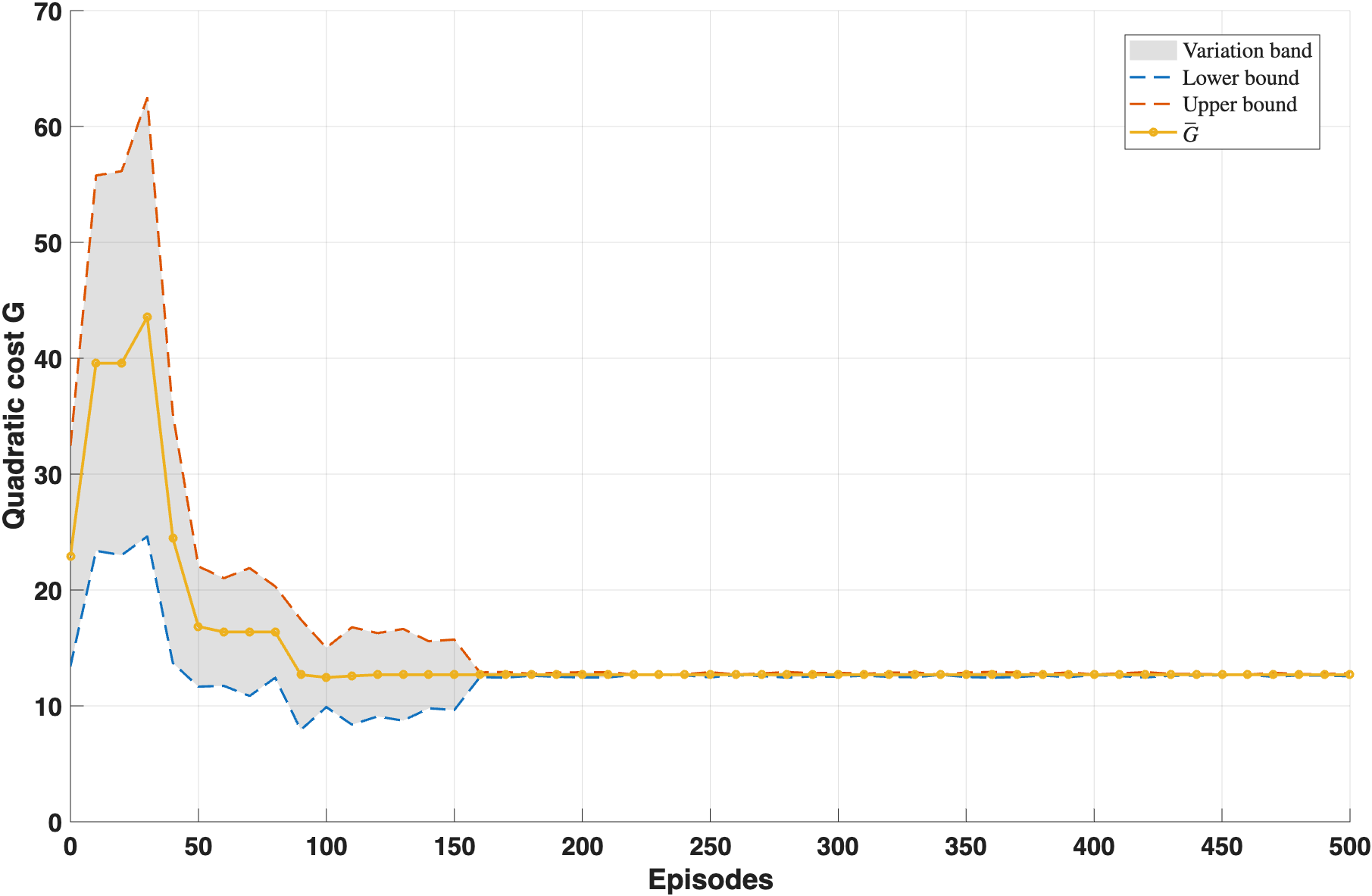}
\caption{Learning stability and convergence for a 6-agent system under ring topology with $\lambda=500$.}
\label{fig:learningconvergance}
\end{figure}

\begin{figure}[!b]
\centering
\includegraphics[width=0.95\linewidth]{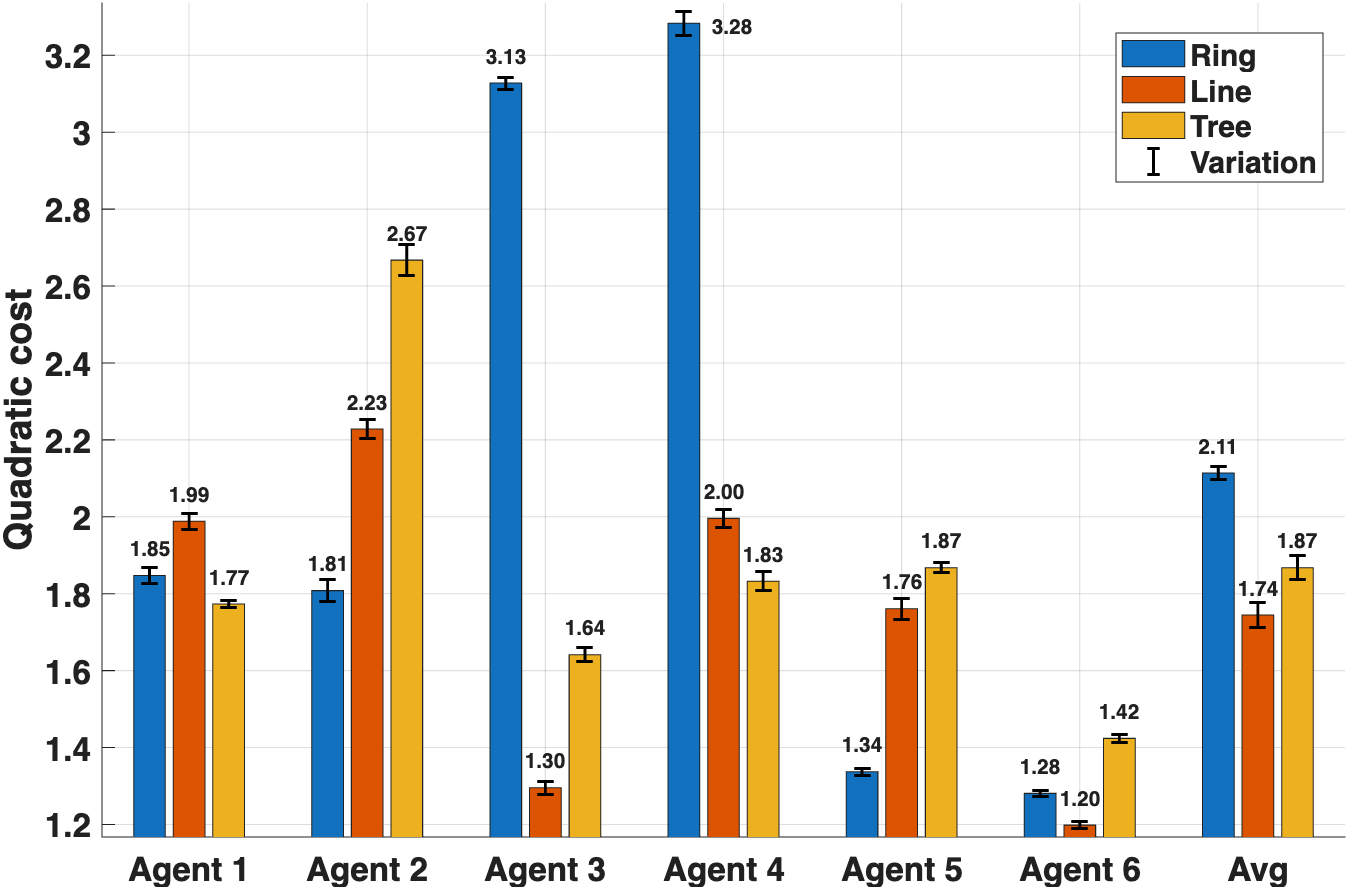}
\caption{Performance distribution among 6-agents of a MAS under different topologies with $\lambda=500$.}
\label{fig:indivitualcost}
\end{figure}

Fig.~\ref{fig:learningconvergance} illustrates the learning stability and convergence behavior of the proposed algorithm for a representative test configuration. The figure shows only the first 500 episodes since the learning converged in earlier episodes. In this experiment, we consider a mid-scale \ac{mas} consisting of six agents in a ring network configuration, where $\lambda=500$ balances the impact of link noise and delay. To ensure learning reliability, the experiment is repeated over five independent random seeds. The solid curve shows the average episode cost $\bar{G}$, calculated for the five runs, while the shaded region depicts $\bar{G} \pm \sigma$, where $\sigma$ denotes the corresponding standard deviation.

\begin{figure}[t]
    \centering
    \begin{subfigure}{0.9\linewidth}
        \centering
        \includegraphics[width=\linewidth]{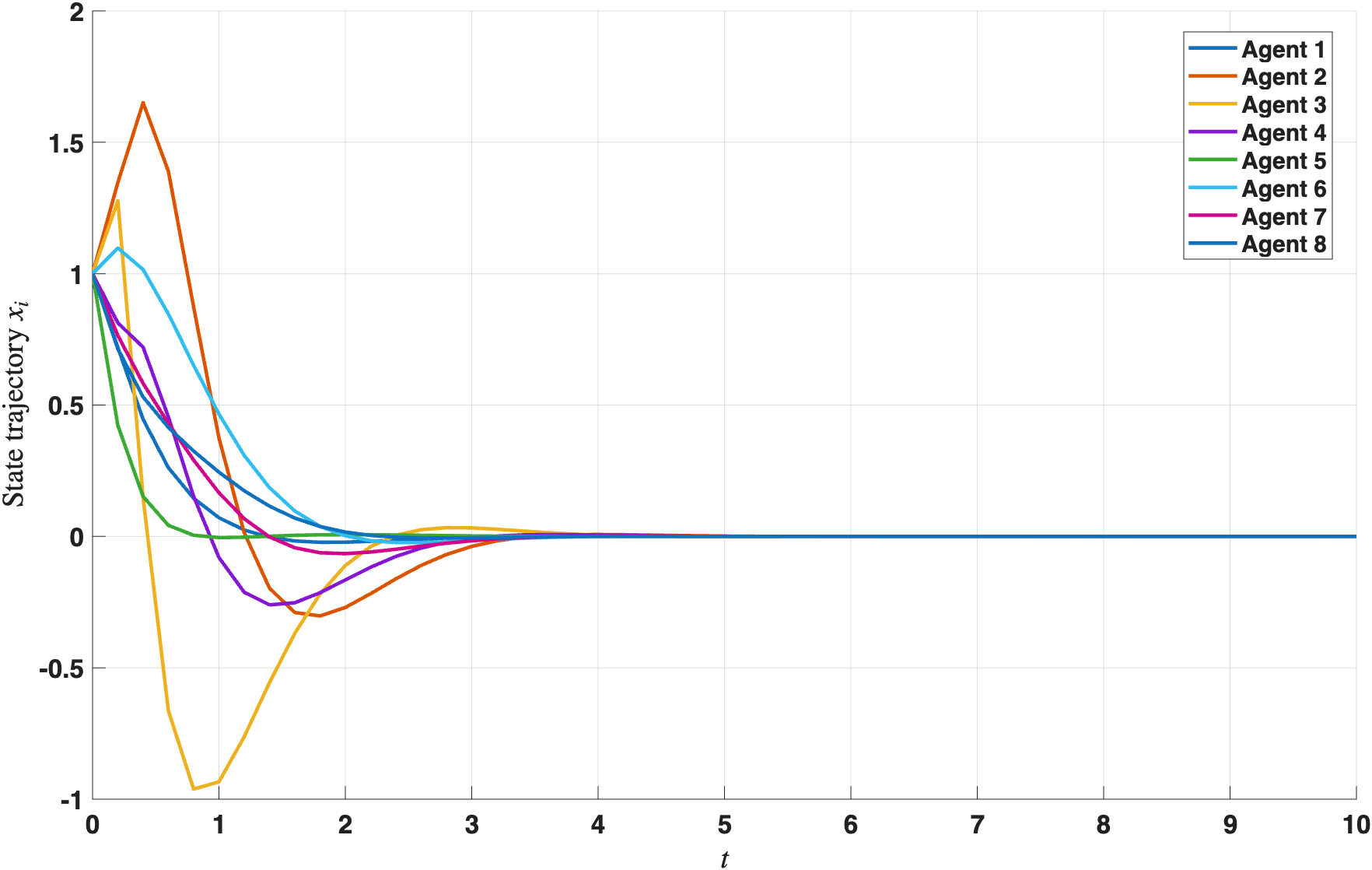}
        \caption{Ring topology}
    \end{subfigure}
\\
    \begin{subfigure}{0.9\linewidth}
        \centering
        \includegraphics[width=\linewidth]{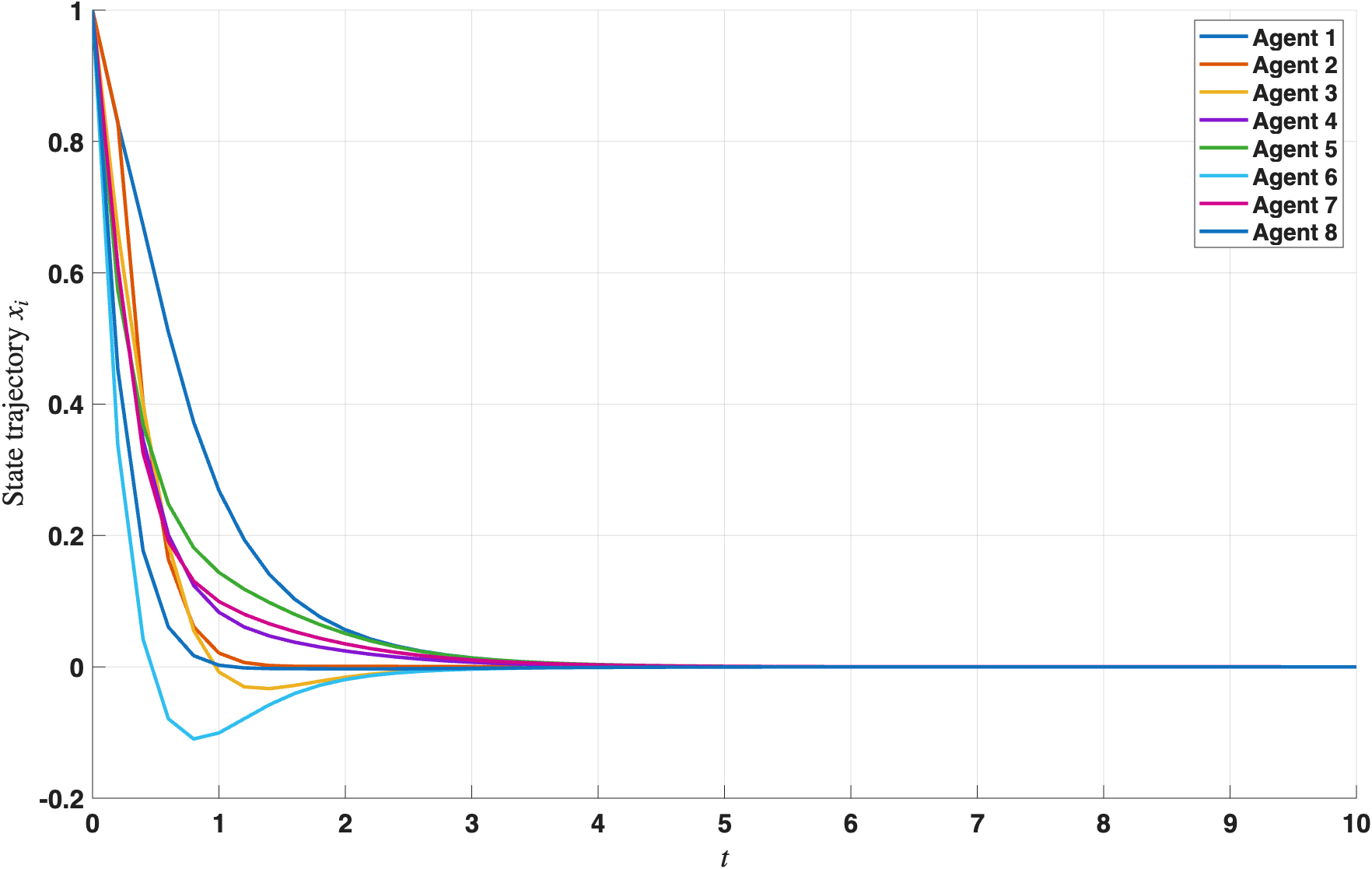}
        \caption{Line topology}
    \end{subfigure}
\\

    \begin{subfigure}{0.9\linewidth}
        \centering
        \includegraphics[width=\linewidth]{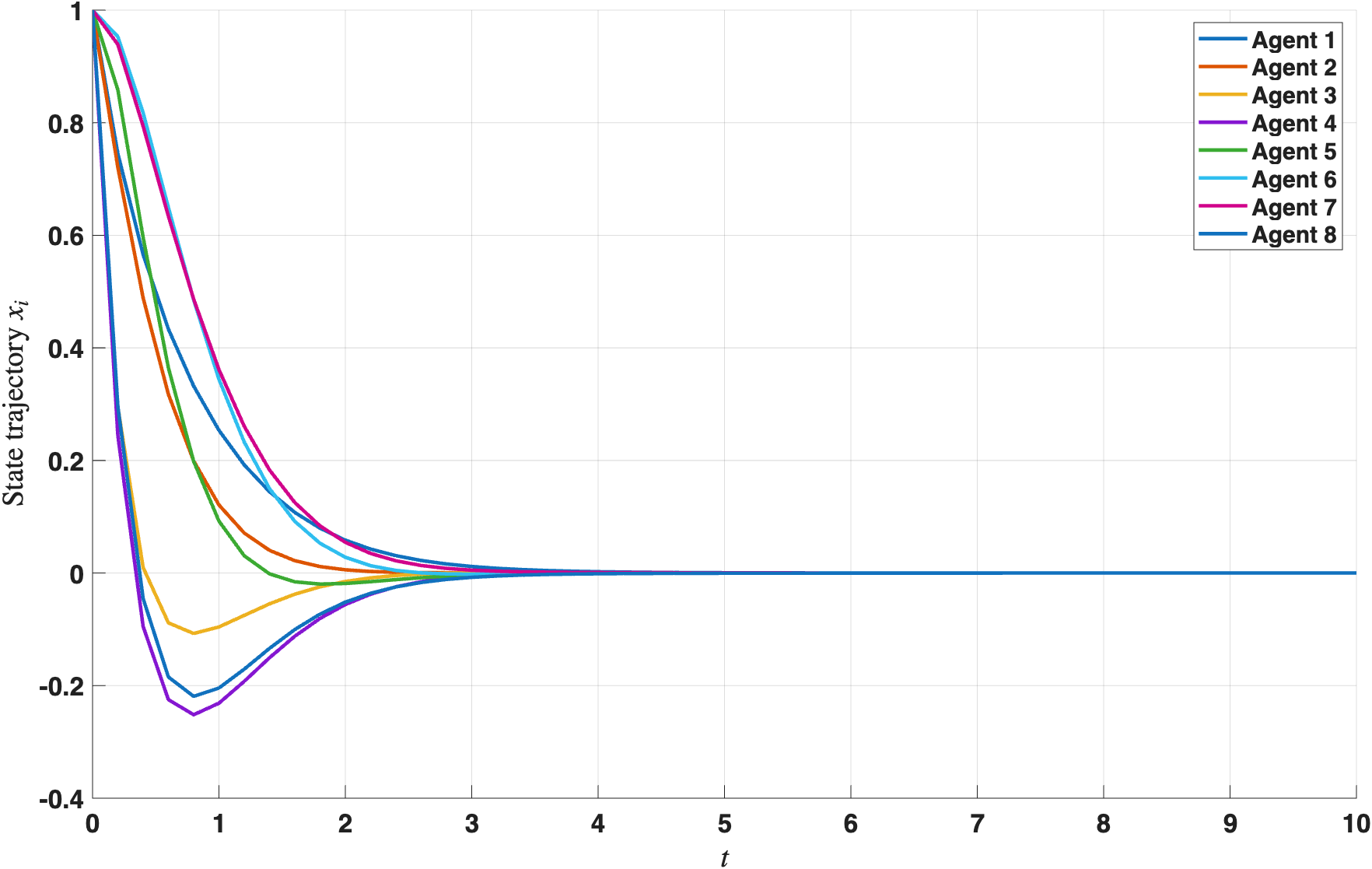}
        \caption{Tree topology}
    \end{subfigure}

    \caption{Trajectories of a MAS with 8 agents}
    \label{fig:traj8}
\end{figure}

The cost~$G$ for each episode initially increases due to exploration, then decreases consistently during learning and stabilizes after a finite number of episodes showing progressive improvement of the learned controller. The cost converges to approximately 
$12.68$ after about 150 episodes. Additionally, the standard deviation across random seeds gradually diminishes and remains within $\pm0.25$ around the mean episode cost, demonstrating stable and repeatable convergence behavior.

To examine the performance distribution, Fig.~\ref{fig:indivitualcost} presents the steady-state quadratic cost for each agent together with the corresponding average cost under three network configurations: line, tree, and ring. For each topology, the reported values are computed by averaging the episode costs over the final 500 learning episodes. Although the steady-state costs vary across agents due to heterogeneous delays, noise levels, and structural positions in the network, all agents exhibit stable convergence to finite cost values. No agent shows divergence, oscillatory behavior, or instability. %The results show that the ring topology yields an average cost of $2.11$ compared to $1.74$ for the line configuration, corresponding to an increase of approximately $21.3\%$. However, this does not imply any advantage of one topology over another, as the noise mean and variance are randomly assigned and the number of links in each topology may be different, as explained before.

To visually validate the closed-loop behavior under the learned controller, Fig.~\ref{fig:traj8} illustrates the state trajectories of all eight agents for a representative test case under the ring, line, and tree network configurations. For each topology, the controller gain corresponding to the best steady-state performance is selected. The trajectories are generated by applying the learned controller to the system dynamics over a ten-step horizon. As shown, the states of all agents remain bounded and converge smoothly toward the equilibrium under all three network configurations. No oscillatory or unstable behavior is observed. The line topology exhibits a smoother transient with lower control effort, whereas the ring topology shows a relatively higher control effort and more overshoot. The maximum observed peak corresponds to a $62\%$ overshoot. Despite these transient deviations, all trajectories settle to the equilibrium in fewer than four time steps across all test cases. Differences in transient response across topologies reflect variations in information flow and connectivity structure.

\begin{figure}[t]
\centering
\includegraphics[width=0.95\linewidth]{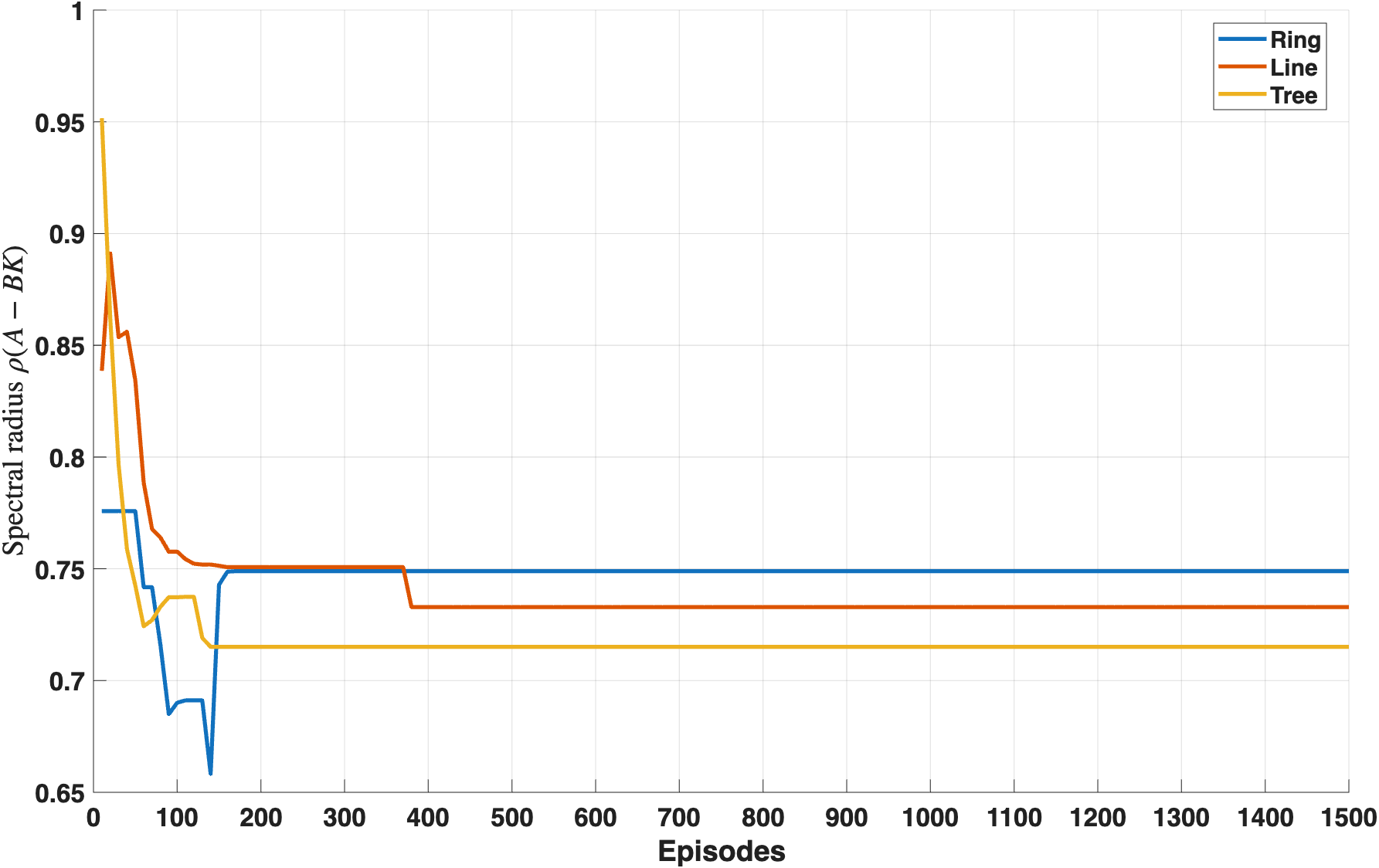}
\caption{Spectral radius of a 5-agent MAS under different topologies with $\lambda=100$.}
\label{fig:spectralrad}
\end{figure}

To further verify closed-loop stability and performance of the learned controller, Fig.~\ref{fig:spectralrad} reports the spectral radius of the resulting closed-loop system matrix for a five-agent network under the line, tree, and ring configurations. After convergence, the learned feedback gain is applied to form the closed-loop matrix. For a discrete-time \ac{lti} system, asymptotic stability is guaranteed if and only if all eigenvalues of the closed-loop system lie strictly inside the unit circle. Accordingly, the spectral radius is defined as
\begin{align}
    \rho(A-BK) = \max |\lambda_i(A-BK)|,
\end{align}
where $\lambda_i(A-BK)$ denotes the $i$th eigenvalue of the closed-loop matrix. As shown in Fig.~\ref{fig:spectralrad}, the spectral radius converges during learning and stabilizes without oscillatory behavior once learning settles. In all topologies, the final spectral radius remains strictly below one and consistently falls within the range of approximately $0.7$ to $0.75$. This indicates that the learned controller achieves a comfortably stable closed-loop system, operating well inside the unit circle rather than near the stability boundary. Such values reflect sufficiently fast state convergence while avoiding excessively aggressive feedback gains.

\subsection{Regret, robustness, and scalability}
This section analyzes the regret, robustness, and scalability of \ac{coco} through simulation studies. The evaluation is based on the cumulative quadratic cost defined in \eqref{eq:masquadraticcost}, which serves as the main performance metric throughout the analysis.

\begin{figure}[t]
\centering
\includegraphics[width=0.95\linewidth]{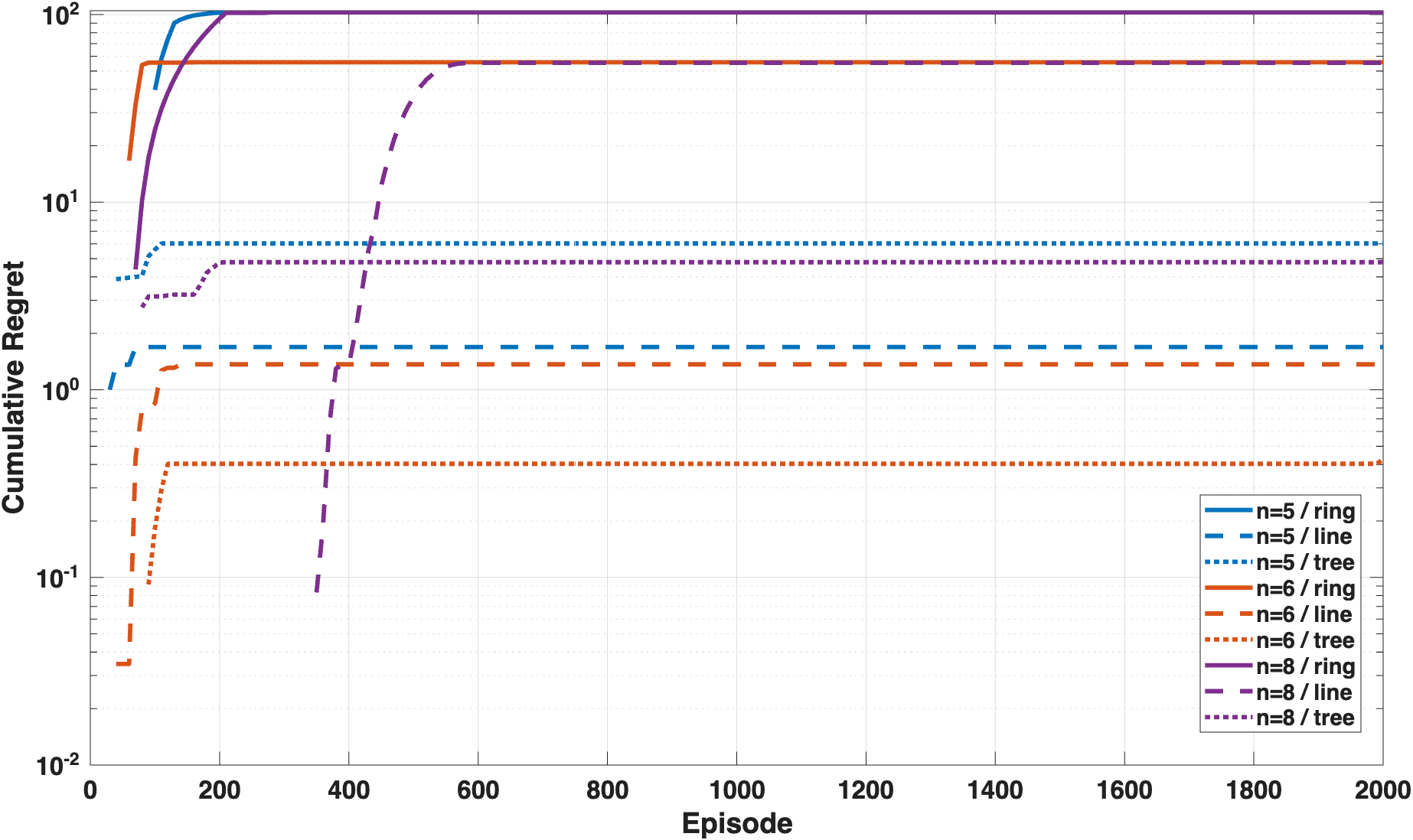}
\caption{Cumulative regret vs. learning episodes. Episodes leading to closed-loop blow-up are removed.}
\label{fig:regret}
\end{figure}

Fig.~\ref{fig:regret} shows the cumulative regret over learning episodes for all agent counts over three network configurations. For each episode~$e$, the collective quadratic cost~$G(e)$ is computed from the closed-loop trajectory. Using the best-so-far reference,
\begin{align}
    G_{\textit{ref}}(e) = \min_{i \leq e} G(i),
\end{align}
the cumulative regret is defined as
\begin{align}
    \textit{Regret}(e) = \sum_{i=0}^{e} G(e) - G_{\textit{ref}}(e).
\end{align}
Episodes that result in closed-loop blow-up are terminated and excluded from the analysis; therefore, some curves begin after the initial episodes. Such early instability is expected during exploration when feedback gains are far from stabilizing values. For the eight-agent case, more initial episodes are excluded, reflecting the increased difficulty of identifying a stabilizing controller in larger networks. In particular, under the line topology, the first stabilizing controller is obtained only after approximately 350 episodes. Nevertheless, all configurations exhibit clear stabilization, i.e., after a transient phase, the regret curves flatten, indicating that learned controllers are close to the best achieved performance. Higher regret values suggest that more exploration is required to find a high-quality controller. This trend is particularly evident for \acp{mas} with the ring topology, where the cumulative regret reaches values on the order of $10^2$.

Additionally, \acp{mas} with fewer agents converge earlier, and the convergence time increases with the network size. For example, systems with eight agents converge in fewer than 600 episodes under the worst-case topology, whereas systems with five agents converge in fewer than 200 episodes under their worst-case topology.

\begin{figure}[t]
\centering
\includegraphics[width=0.95\linewidth]{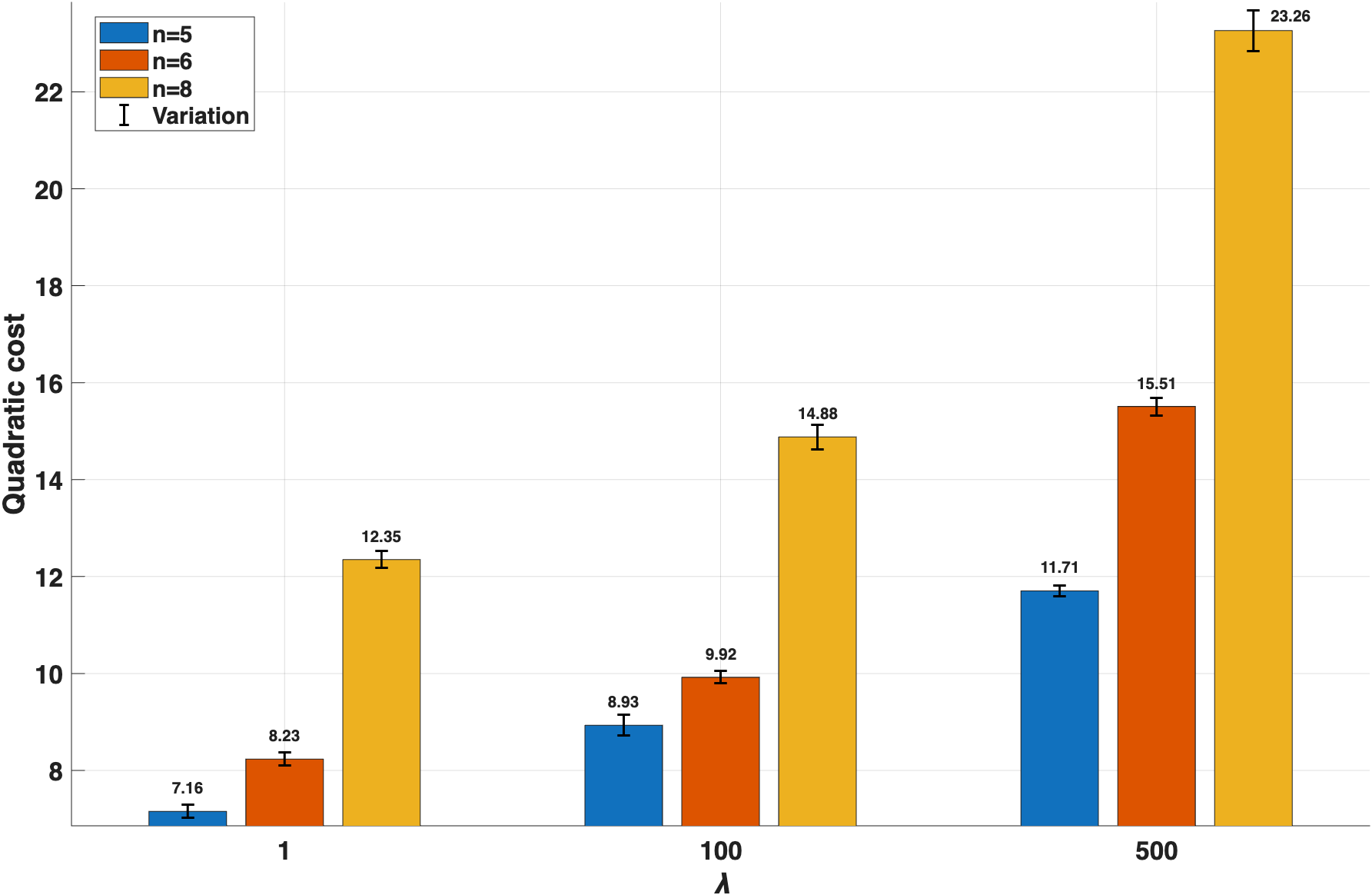}
\caption{Robustness of CDNet with respect to the delay–noise tradeoff parameter~$\lambda$. The final quadratic cost is reported for $n=5, 6,$ and $8$ agents under a connected degree-3 network topology. Error bars represent the variability across independent runs.}
\label{fig:robust}
\end{figure}

Fig.~\ref{fig:robust} illustrates the effect of the delay–noise balancing parameter~$\lambda \in \{1,100,500\}$ on the final collective quadratic cost~$G$ for systems with~$n=5,6,$ and $8$ agents. To highlight robustness, we consider a connected network topology with a node degree of three. This moderately dense topology introduces multiple routing alternatives per node which amplifies the influence of the delay–noise weighting on the selected communication paths. The reported values correspond to the average cost over the last 1000 learning episodes, ensuring that the results reflect steady-state behavior rather than transient learning effects. The error bars represent the variability of four independent random seeds. Across all tested configurations, the relatively small variation confirms stable convergence under different tradeoff regimes. 

As depicted in the figure, the cost increases monotonically with $\lambda$ for all network sizes. When $\lambda=1$, delay dominates the routing objective, leading to shorter communication paths and lower overall control cost. In this case, the final costs for the~$n=5,6,$ and $8$ agent test cases are $7.16$, $8.23$, and $12.35$, respectively. As $\lambda$ increases, the routing mechanism prioritizes lower-noise links even at the expense of additional hops. This leads to a substantial increase in cost, ranging from $64\%$ for the 5-agent \ac{mas} to $91\%$ for the 8-agent \ac{mas}, with an increase of $89\%$ observed in the case of 6 agents.

\begin{figure}[t]
\centering
\includegraphics[width=0.95\linewidth]{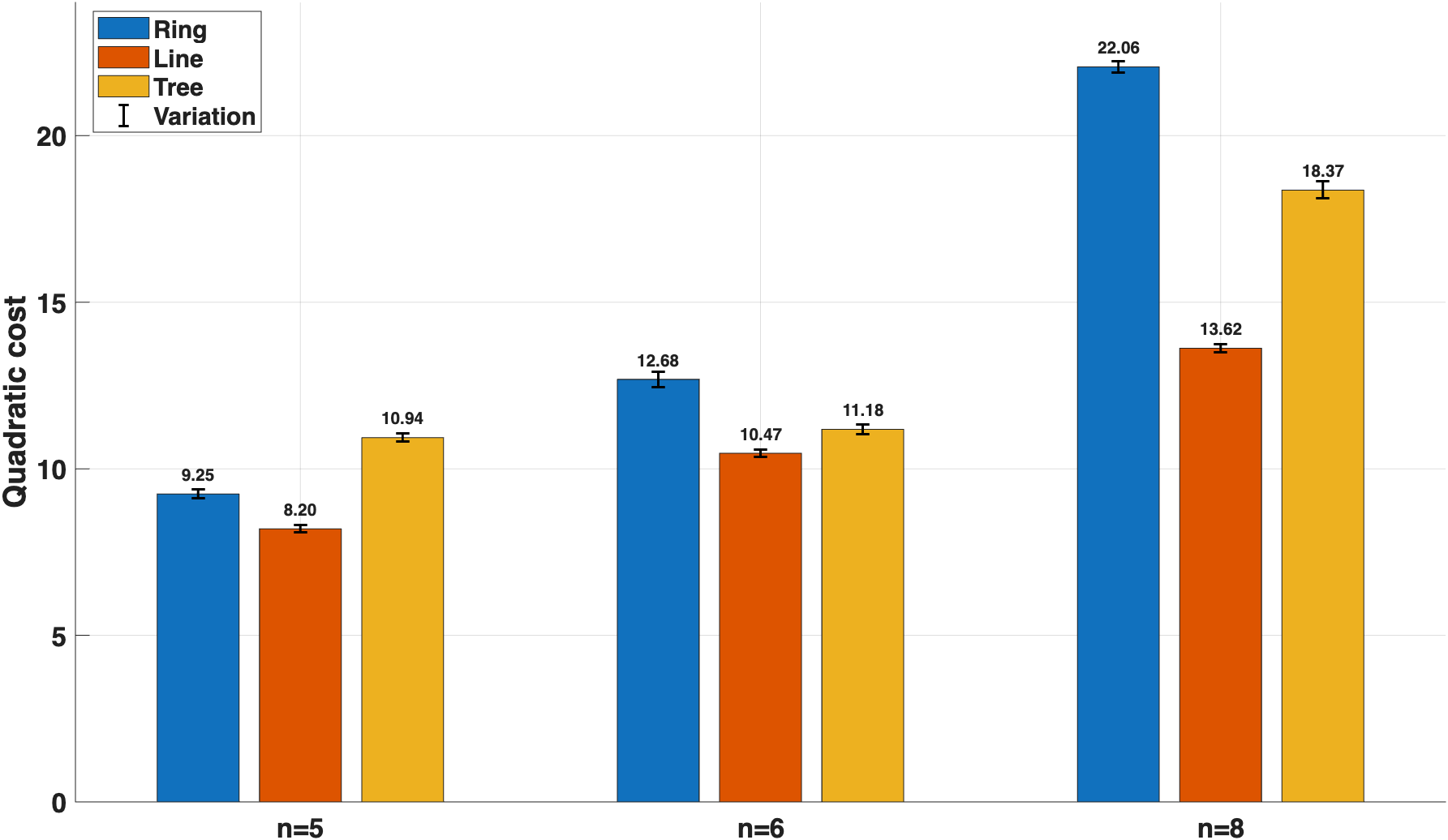}
\caption{Scalability of CDNet with MAS size.}
\label{fig:scale}
\end{figure}

Fig.~\ref{fig:scale} shows the quadratic cost for \acp{mas} with~$n=5,6,$ and $8$ agents under ring, line, and tree topologies with the same noise parameters. The reported values are averaged over the last 1000 learning episodes, and the error bars represent variability across these episodes. As the \ac{mas} size increases, the quadratic cost increases across all topologies, reflecting the higher-dimensional state space and increased complexities. The line topology consistently yields the lowest cost. Relative to the 5-agent baseline cost of $8.20$, the cost increases by $28\%$ for 6 agents and $66\%$ for 8 agents. The second best topology is the tree topology, with a baseline cost of $10.94$ for 5 agents and increases of $2\%$ and $68\%$ for 6 and 8 agents, respectively. In contrast, the ring topology exhibits the highest cost, reaching $22.06$ for the 8-agent \ac{mas} corresponding to a $138\%$ increase compared to the 5-agent case. In all configurations, learning converged and stabilized within at most 800 episodes. The small variability confirms stable learning dynamics even for the largest \ac{mas}.

\subsection{Comparison with related work}
In this section, we briefly present the following baseline solutions and compare their performance with \ac{coco}: 
\subsubsection{Analytic Riccati solution (OPT)}
In this approach, the optimal control is calculated using the discrete time algebraic Riccati equation as discussed in Section~\ref{sec:preliminaries}.

\subsubsection{Distributed state tracking (DST)}
This approach is proposed in~\cite{wang2020distributed} and synthesizes individual control for each agent using the estimation of the global state. This approach ignores link noise and delay. We update the weighting coefficients for immediate state estimation to capture the network configurations. 
In this approach, each agent defines a Q-function that captures the cumulative quadratic control cost using its individual observation as
\begin{align}
    \label{eq:q-function}
    Q_\ell(\bm{X}_\ell(t), \bm{u}_\ell(t)) = c_\ell + Q_\ell(\bm{X}_\ell(t+1), \bm{u}_\ell(t+1)),
\end{align}
where $c_\ell = \bm{X}_\ell(t)^{\top} S_\ell \bm{X}_\ell(t) + \bm{u}_\ell(t)^{\top} R_\ell \bm{u}_\ell(t)$ and $\bm{u}_\ell(t) = K_\ell\bm{Z}_\ell(t)$. Thus, the individual Q-functions depend only on each agent's state and individual observation. On the one hand, this Q-function can be rewritten in an expanded quadratic form of~\cite{wang2020distributed}
\begin{align}
    \label{eq:statetrackingquadratic}
    Q_\ell(\bm{X}_\ell(t), \bm{u}_\ell(t)) = \left[\bm{Z}_\ell(t);\bm{u}_\ell(t)\right]^{\top} \mathcal{H}_\ell \left[\bm{Z}_\ell(t);\bm{u}_\ell(t)\right],
\end{align}
where $\mathcal{H}_\ell$ is a symmetric block matrix of $2 \times 2$ blocks. On the other hand, as shown in \cite{wang2020distributed}, the Q-function can be approximated with the individual estimations~$\bm{Z}_\ell$ as
\begin{align}
\label{eq:statetrackingQapprox}
    Q_\ell(\bm{X}_\ell(t), \bm{u}_\ell(t)) \approx \bm{y}_\ell(t) \bm{\theta}_\ell, 
\end{align}
where $y_\ell(t)$ is a vector that contains all quadratic bases over the elements in $\left[\bm{Z}_\ell(t);\bm{u}_\ell(t)\right]$ and $\bm{\theta}_\ell$ is the coefficient vector. This vector is calculated as $y_\ell(t) = \left[\bm{z}_1^2(t), \bm{z}_1(t)\bm{z}_2(t),\cdots, \bm{z}_L(t)\bm{u}_\ell(t), \bm{u}_\ell^2(t)\right]$.

Since $y_\ell(t)$ and $y_\ell(t+1)$ are known, using \eqref{eq:q-function} and \eqref{eq:statetrackingQapprox}, the coefficient factor is learned using online gradient descent methods to minimize the loss
\begin{align}
    \sum_{t=0}^{N}||\left( \bm{y}_\ell(t) - \bm{y}_\ell(t+1)\right) \bm{\theta}_\ell -c_\ell||^2
\end{align}
over $t \in N$ evaluation time steps~\cite{wang2020distributed}. Then, using \eqref{eq:statetrackingquadratic} and \eqref{eq:statetrackingQapprox}, the matrix $\mathcal{H}_\ell$ is determined. Finally, the optimal control~$K_\ell^*$ is calculated as
\begin{align}
    K_\ell^* = - \mathcal{H}_{\ell,22}^{-1}\mathcal{H}_{\ell,21},
\end{align}
where $\mathcal{H}_{\ell,22}$ and $\mathcal{H}_{\ell,21}$ are the blocks in $\mathcal{H}_\ell$.

\begin{figure}[t]
\centering
\includegraphics[width=0.95\linewidth]{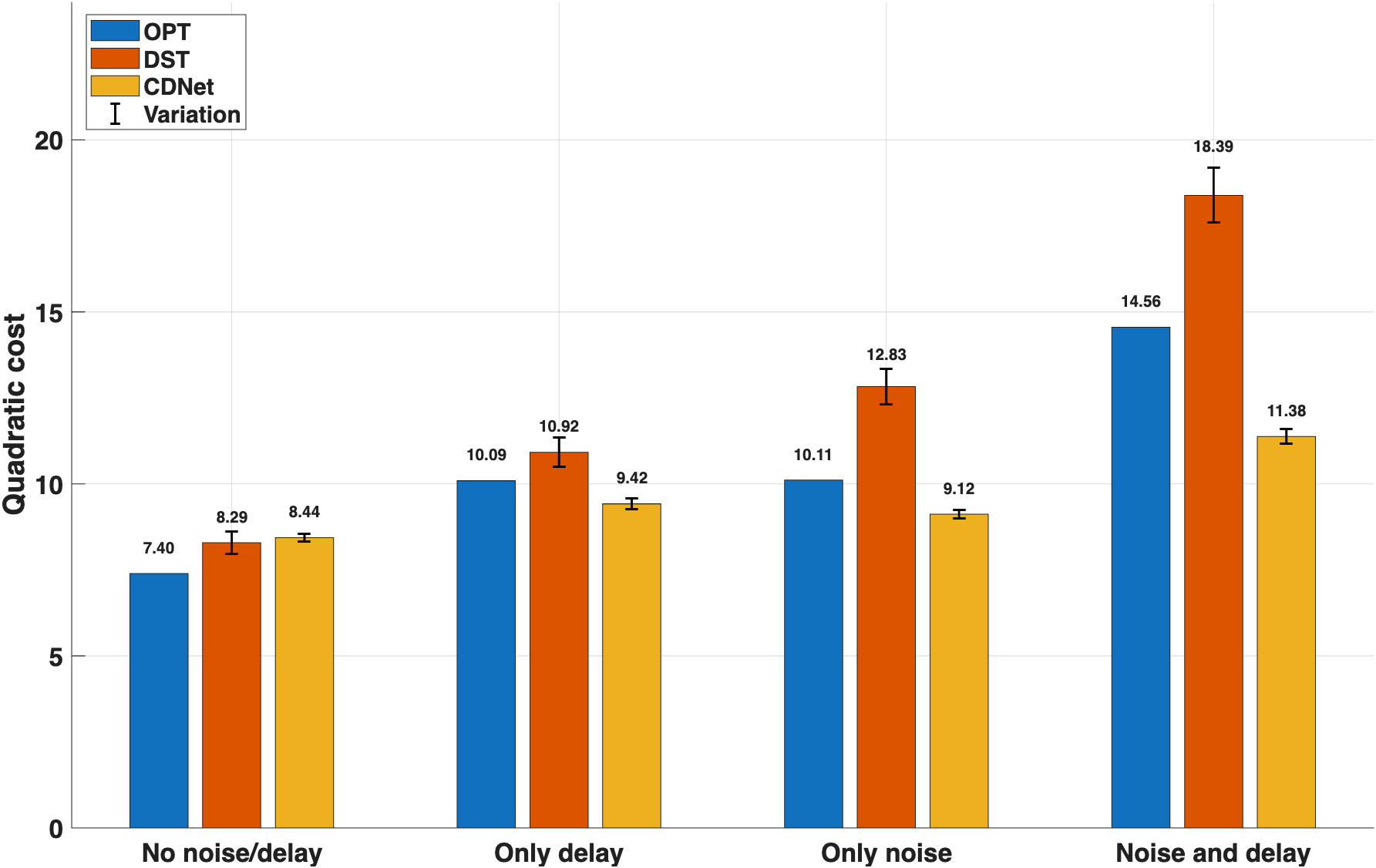}
\caption{Comparing CDNet with the related work.}
\label{fig:baseline}
\end{figure}

We compare the performance of \ac{coco} with those of the baseline methods in Fig.~\ref{fig:baseline}. The evaluation considers four communication scenarios: (\emph{i}) ideal communication (no delay and no noise), (\emph{ii}) communication delay only, (\emph{iii}) link noise only, and (\emph{iv}) simultaneous delay and noise. The reported quadratic costs correspond to the collective cost of a six-agent \ac{mas} under a ring topology that is calculated for a roll-out of 20 steps. For simplicity, we have only considered $N(0, 0.02)$ as the noise model. This is because the other approaches will diverge since they do not have a noise compensation mechanism.
The optimal controller (OPT) represents the centralized analytic solution under perfect communication; therefore, no variation is reported for OPT and only the average value is reported whenever needed. The results for DST and \ac{coco} are averaged over five independent random seeds, and the error bars indicate the corresponding variability of different runs.

As shown, under ideal communication, all methods achieve comparable performance, with OPT providing the lowest cost of $7.40$ as expected. In the presence of delay or noise individually, \ac{coco} outperforms the other methods, while DST exhibits the largest degradation. In particular, OPT and DST experience the maximum cost increases of $55\%$ and $37\%$, respectively, while \ac{coco} shows only a degradation of $12\%$. The results further indicate that \ac{coco} is more sensitive to delay than to noise, as noise alone increases the cost by only $8\%$. The performance gap becomes larger when both delay and noise are present.  In this scenario, \ac{coco}  maintains a substantially lower cost, demonstrating improved robustness. DST performs the worst, with $122\%$ increase in cost. Note that all links are assumed to have identical noise statistics with zero bias; otherwise, OPT and DST may lose closed-loop stability.

\section{Conclusion}
\label{sec:conclusion}
This paper presented an encoded double deep Q-network reinforcement learning framework for distributed control of multi-agent systems. The objective was to determine individual control policies that collectively minimize the quadratic cost of all agents. Information exchange among agents occurs over a communication network subject to delay, sampling asynchrony, and noise. A multi-hop routing algorithm is designed to balance the effects of delay and noise. The proposed approach integrates a message-passing mechanism that leverages the network topology to model observation noise, refine global state estimates, align agents’ relative clocks, and time-shift the global state information. As time-shifted estimates become available, each agent updates its Q-networks asynchronously to improve the learned policy. To mitigate overestimation bias caused by noisy observations, a pessimistic Q-value selection strategy is employed. The proposed method was evaluated on multiple test cases and scenarios. Future work will focus on extending the framework to dynamic and time-varying network configurations.

\balance
\bibliography{ref}

\begin{thebibliography}{10}

\bibitem{whatisMAS}
L.~Busoniu, R.~Babuska, and B.~De~Schutter, ``A comprehensive survey of multiagent reinforcement learning,'' {\em IEEE Transactions on Systems, Man, and Cybernetics, Part C (Applications and Reviews)}, vol.~38, no.~2, pp.~156--172, 2008.

\bibitem{featureMAS}
H.~Pei, ``Group consensus of multi-agent systems with hybrid characteristics and directed topological networks,'' {\em ISA Transactions}, vol.~138, pp.~311--317, 2023.

\bibitem{alemzadeh2019distributed}
S.~Alemzadeh and M.~Mesbahi, ``Distributed {Q-learning} for dynamically decoupled systems,'' in {\em American Control Conference (ACC)}, pp.~772--777, IEEE, 2019.

\bibitem{wang2020distributed}
H.~Wang, S.~Lin, H.~Jafarkhani, and J.~Zhang, ``Distributed {Q-Learning} with state tracking for multi-agent networked control,'' in {\em Proceedings of the 20th International Conference on Autonomous Agents and MultiAgent Systems}, AAMAS '21, (Richland, SC), p.~1692–1694, International Foundation for Autonomous Agents and Multiagent Systems, 2021.

\bibitem{d2003distributed}
R.~D'Andrea and G.~E. Dullerud, ``Distributed control design for spatially interconnected systems,'' {\em IEEE Transactions on Automatic Control}, vol.~48, no.~9, pp.~1478--1495, 2003.

\bibitem{massioni2009distributed}
P.~Massioni and M.~Verhaegen, ``Distributed control for identical dynamically coupled systems: A decomposition approach,'' {\em IEEE Transactions on Automatic Control}, vol.~54, no.~1, pp.~124--135, 2009.

\bibitem{hoffmann2013distributed}
C.~Hoffmann, A.~Eichler, and H.~Werner, ``Distributed control of linear parameter-varying decomposable systems,'' in {\em American Control Conference}, pp.~2380--2385, IEEE, 2013.

\bibitem{venkat2005stability}
A.~N. Venkat, J.~B. Rawlings, and S.~J. Wright, ``Stability and optimality of distributed model predictive control,'' in {\em Proceedings of the IEEE Conference on Decision and Control}, pp.~6680--6685, IEEE, 2005.

\bibitem{dunbar2007distributed}
W.~B. Dunbar, ``Distributed receding horizon control of dynamically coupled nonlinear systems,'' {\em IEEE Transactions on Automatic Control}, vol.~52, no.~7, pp.~1249--1263, 2007.

\bibitem{li2015fully}
Z.~Li and Z.~Ding, ``Fully distributed adaptive consensus control of multi-agent systems with {LQR} performance index,'' in {\em 2015 54th IEEE Conference on Decision and Control (CDC)}, pp.~386--391, IEEE, 2015.

\bibitem{chang2023regret}
T.-J. Chang and S.~Shahrampour, ``Regret analysis of distributed online {LQR} control for unknown {LTI} systems,'' {\em IEEE Transactions on Automatic Control}, vol.~69, no.~1, pp.~667--673, 2023.

\bibitem{borrelli2008distributed}
F.~Borrelli and T.~Keviczky, ``Distributed {LQR} design for identical dynamically decoupled systems,'' {\em IEEE Transactions on Automatic Control}, vol.~53, no.~8, pp.~1901--1912, 2008.

\bibitem{vlahakis2019distributed}
E.~E. Vlahakis, L.~D. Dritsas, and G.~D. Halikias, ``Distributed {LQR} design for identical dynamically coupled systems: Application to load frequency control of multi-area power grid,'' in {\em 2019 IEEE 58th Conference on Decision and Control (CDC)}, pp.~4471--4476, IEEE, 2019.

\bibitem{dong2010distributed}
W.~Dong, ``Distributed optimal control of multiple systems,'' {\em International Journal of Control}, vol.~83, no.~10, pp.~2067--2079, 2010.

\bibitem{watkins1992q}
C.~J. Watkins and P.~Dayan, ``Q-learning,'' {\em Machine learning}, vol.~8, pp.~279--292, 1992.

\bibitem{narayanan2016distributed}
V.~Narayanan and S.~Jagannathan, ``Distributed adaptive optimal regulation of uncertain large-scale interconnected systems using hybrid {Q-learning} approach,'' {\em IET Control Theory \& Applications}, vol.~10, no.~12, pp.~1448--1457, 2016.

\bibitem{dizche2019sparse}
A.~F. Dizche, A.~Chakrabortty, and A.~Duel-Hallen, ``Sparse wide-area control of power systems using data-driven reinforcement learning,'' in {\em 2019 American Control Conference (ACC)}, pp.~2867--2872, IEEE, 2019.

\bibitem{zhang2018fully}
K.~Zhang, Z.~Yang, H.~Liu, T.~Zhang, and T.~Basar, ``Fully decentralized multi-agent reinforcement learning with networked agents,'' in {\em International Conference on Machine Learning}, pp.~5872--5881, PMLR, 2018.

\bibitem{zhang2019distributed}
Y.~Zhang and M.~M. Zavlanos, ``Distributed off-policy actor-critic reinforcement learning with policy consensus,'' in {\em IEEE Conference on Decision and Control}, pp.~4674--4679, IEEE, 2019.

\bibitem{ma2024efficient}
C.~Ma, A.~Li, Y.~Du, H.~Dong, and Y.~Yang, ``Efficient and scalable reinforcement learning for large-scale network control,'' {\em Nature Machine Intelligence}, vol.~6, no.~9, pp.~1006--1020, 2024.

\bibitem{kayaalp2023policy}
M.~Kayaalp, F.~Ghadieh, and A.~H. Sayed, ``Policy evaluation in decentralized {POMDPS} with belief sharing,'' {\em IEEE Open Journal of Control Systems}, vol.~2, pp.~125--145, 2023.

\bibitem{sharedmemory2020}
W.~Mao, K.~Zhang, E.~Miehling, and T.~Başar, ``Information state embedding in partially observable cooperative multi-agent reinforcement learning,'' in {\em 2020 59th IEEE Conference on Decision and Control (CDC)}, pp.~6124--6131, 2020.

\bibitem{offlinebeliefgeneration2021}
P.~Moreno, E.~Hughes, K.~R. McKee, B.~A. Pires, and T.~Weber, ``Neural recursive belief states in multi-agent reinforcement learning,'' 2021.
\newblock Preprint. Available at arXiv:2102.02274.

\bibitem{he2019adaptive}
S.~He, H.~Fang, M.~Zhang, F.~Liu, and Z.~Ding, ``Adaptive optimal control for a class of nonlinear systems: The online policy iteration approach,'' {\em IEEE Transactions on Neural Networks and Learning Systems}, vol.~31, no.~2, pp.~549--558, 2019.

\bibitem{aysal2008distributed}
T.~C. Aysal, M.~J. Coates, and M.~G. Rabbat, ``Distributed average consensus with dithered quantization,'' {\em IEEE Transactions on Signal Processing}, vol.~56, no.~10, pp.~4905--4918, 2008.

\bibitem{kar2008distributed}
S.~Kar and J.~M. Moura, ``Distributed consensus algorithms in sensor networks with imperfect communication: Link failures and channel noise,'' {\em IEEE Transactions on Signal Processing}, vol.~57, no.~1, pp.~355--369, 2008.

\bibitem{rajagopal2010network}
R.~Rajagopal and M.~J. Wainwright, ``Network-based consensus averaging with general noisy channels,'' {\em IEEE Transactions on Signal Processing}, vol.~59, no.~1, pp.~373--385, 2010.

\bibitem{griparic2022consensus}
K.~Griparic, M.~Polic, M.~Krizmancic, and S.~Bogdan, ``Consensus-based distributed connectivity control in multi-agent systems,'' {\em IEEE Transactions on Network Science and Engineering}, vol.~9, no.~3, pp.~1264--1281, 2022.

\bibitem{ortega2024quantized}
T.~Ortega and H.~Jafarkhani, ``Quantized and asynchronous federated learning,'' {\em IEEE Transactions on Communications}, vol.~73, no.~4, pp.~2361--2374, 2024.

\bibitem{ortega2024decentralized}
T.~Ortega and H.~Jafarkhani, ``Decentralized optimization in networks with arbitrary delays,'' in {\em ICC 2024-IEEE International Conference on Communications}, pp.~794--799, IEEE, 2024.

\bibitem{shen2021distributed}
Y.~Shen, S.~Karimi-Bidhendi, and H.~Jafarkhani, ``Distributed and quantized online multi-kernel learning,'' {\em IEEE Transactions on Signal Processing}, vol.~69, pp.~5496--5511, 2021.

\bibitem{diaz2025multi}
C.~Diaz-Vilor, M.~Barzegaran, and H.~Jafarkhani, ``Multi-{UAV} energy-efficient wildfire coverage optimization,'' {\em IEEE Transactions on Wireless Communications}, 2025.

\bibitem{barzegaran2025dynamic}
M.~Barzegaran and H.~Jafarkhani, ``Dynamic deployment of heterogeneous wireless sensor drone networks with limited communication range,'' {\em IEEE Transactions on Vehicular Technology}, 2025.

\bibitem{tung2021effective}
T.-Y. Tung, S.~Kobus, J.~P. Roig, and D.~G{\"u}nd{\"u}z, ``Effective communications: A joint learning and communication framework for multi-agent reinforcement learning over noisy channels,'' {\em IEEE Journal on Selected Areas in Communications}, vol.~39, no.~8, pp.~2590--2603, 2021.

\bibitem{yang2023collision}
Y.~Yang, Q.~Liu, H.~Tan, Z.~Shen, and D.~Wu, ``Collision-free and connectivity-preserving formation control of nonlinear multi-agent systems with external disturbances,'' {\em IEEE Transactions on Vehicular Technology}, vol.~72, no.~8, pp.~9956--9968, 2023.

\bibitem{9930941}
C.~Diaz-Vilor, A.~Lozano, and H.~Jafarkhani, ``Cell-free {UAV} networks: Asymptotic analysis and deployment optimization,'' {\em IEEE Transactions on Wireless Communications}, vol.~22, no.~5, pp.~3055--3070, 2023.

\bibitem{10186347}
C.~Diaz-Vilor, A.~Lozano, and H.~Jafarkhani, ``Cell-free {UAV} networks with wireless fronthaul: Analysis and optimization,'' {\em IEEE Transactions on Wireless Communications}, vol.~23, no.~3, pp.~2054--2069, 2024.

\bibitem{overestimate}
S.~Thrun and A.~Schwartz, ``Issues in using function approximation for reinforcement learning,'' in {\em Proceedings of the 4th Connectionist Models Summer School}, Lawrence Erlbaum Associates, 1993.

\bibitem{shamma2008cooperative}
J.~Shamma, {\em Cooperative control of distributed multi-agent systems}.
\newblock John Wiley \& Sons, 2008.

\bibitem{de2006decentralized}
M.~C. De~Gennaro and A.~Jadbabaie, ``Decentralized control of connectivity for multi-agent systems,'' in {\em Proceedings of the IEEE Conference on Decision and Control}, pp.~3628--3633, IEEE, 2006.

\bibitem{ding2019survey}
D.~Ding, Q.-L. Han, Z.~Wang, and X.~Ge, ``A survey on model-based distributed control and filtering for industrial cyber-physical systems,'' {\em IEEE Transactions on Industrial Informatics}, vol.~15, no.~5, pp.~2483--2499, 2019.

\bibitem{ogata2002modern}
K.~Ogata and Y.~Yang, {\em Modern control engineering}, vol.~4.
\newblock Prentice Hall India, 2002.

\bibitem{ISO_GUM_2008}
{International Organization for Standardization}, ``Evaluation of measurement data — guide to the expression of uncertainty in measurement ({GUM}),'' tech. rep., ISO/IEC Guide 98-3, 2008.

\bibitem{shomorony2013worst}
I.~Shomorony and A.~S. Avestimehr, ``Worst-case additive noise in wireless networks,'' {\em IEEE Transactions on Information Theory}, vol.~59, no.~6, pp.~3833--3847, 2013.

\bibitem{gbadamosi2020design}
O.~A. Gbadamosi and D.~R. Aremu, ``Design of a modified {Dijkstra’s} algorithm for finding alternate routes for shortest-path problems with huge costs,'' in {\em 2020 International Conference in Mathematics, Computer Engineering and Computer Science (ICMCECS)}, pp.~1--6, IEEE, 2020.

\bibitem{bishop2001introduction}
G.~Bishop, G.~Welch, {\em et~al.}, ``An introduction to the {Kalman} filter,'' {\em Proc of SIGGRAPH, Course}, vol.~8, no.~27599-23175, p.~41, 2001.

\end{thebibliography}
\bibliographystyle{ieeetr}

\end{document}